# Interleaved bond frustration in a triangular lattice antiferromagnet

S. J. Gomez Alvarado,[1,†] J. R. Chamorro,[1,†] D. Rout,[1] J. Hielscher,[1,2] Sarah Schwarz,[1] Caeli Benyacko[1], M. B. Stone,[3] V. Ovidiu Garlea,[3] A. R. Jackson,[1] G. Pokharel,[1] R. Gomez,[1] B. R. Ortiz,[1,3,4] Suchismita Sarker,[5] L. Kautzsch,[1] L. C. Gallington,[6] R. Seshadri,[1,7] and Stephen D. Wilson[1,*]

[1]*Materials Department, University of California, Santa Barbara, CA, USA*

[2]*Department of Physics, Technical University of Munich, Munich, Germany*

[3]*Neutron Scattering Division, Oak Ridge National Laboratory, Oak Ridge, TN, USA*

[4]*Materials Science and Technology Division, Oak Ridge National Laboratory, Oak Ridge, TN, USA*

[5]*Cornell High Energy Synchrotron Source, Cornell University, Ithaca, NY, USA*

[6]*X-ray Science Division, Advanced Photon Source, Argonne National Laboratory, Argonne, IL, USA*

[7]*Materials Research Laboratory, University of California, Santa Barbara, CA, USA*

[†]*These authors contributed equally to this work.*

*Email: stephendwilson@ucsb.edu

## Abstract

Frustration of long-range order via lattice geometries serves to amplify fluctuations of the order parameter and generate unconventional ground states that are highly sensitive to perturbations. Traditionally, this concept of geometric frustration is used to engineer unconventional magnetic states in a variety of materials; however, the charge degree of freedom and bond order can be similarly frustrated. Finding materials that host both frustrated magnetic and bond networks holds promise for engineering structural and magnetic states with the potential of coupling to one another via either the magnetic sector (via magnetic field) or via the lattice sector (via strain). In this paper, we identify an unusual instance of this coexistence in the triangular lattice antiferromagnetic compounds $Ln$Cd$_3$P$_3$ ($Ln$ = La, Ce, Pr, and Nd). These compounds feature two-dimensional planes of unique trigonal-planar CdP$_3$ units that manifest an underlying bond instability with its long-range ordering frustrated via emergent kagome ice bond correlations. Our results establish $Ln$Cd$_3$P$_3$ as a rare class of materials where frustrated magnetism across a tunable rare-earth triangular network is embedded within a dopable semiconductor with a frustrated bond order instability.

I.   INTRODUCTION

Geometrical frustration on triangular lattice networks can lead to a number of unconventional states arising from the competition between electronic degrees of freedom. Magnetic interactions, for example, across the triangular lattice network are predicted to stabilize states such as the quantum spin liquids or native quantum disordered ground states.[1–12] This frustration is, however, not limited to spin interactions, and the charge degree of freedom can also be impacted via forms of orbital or bond frustration.[13–18]

Interfacing the charge degree of freedom with highly frustrated magnetic states is an experimental challenge in the field of quantum materials. Creating such an interface is key to testing many predictions of unconventional superconductivity, quantum criticality, and other phenomena that are predicted to arise when itinerant charge carriers interact within a fluctuating background of entangled spins.[19–26] However, many of the leading materials candidates hosting a highly frustrated magnetic lattice are built from highly localized lanthanide ($Ln$) ions whose charge degree of freedom is strongly gapped out and are not readily dopable.[1] Conversely, many dopable magnetic materials are often built from transition metal ions that possess extended exchange interactions or other interactions that lower crystallographic symmetries and lift magnetic frustration.[27–29] Bridging this divide and establishing a cleanly tunable material platform that allows for conduction electrons to coexist within a frustrated, local moment lattice network remains an important goal.

There has been recent progress in stabilizing quantum disordered or quantum spin liquid states across triangular lattice networks of $S_{eff} = 1/2$ magnetic moments derived from lanthanide ($Ln$) ions. A number of quantum disordered or nearly disordered states have been reported in layered $ALnX_2$ ($A$=alkali metal, $X$=O, S, Se, Te) compounds, for instance;[3,30–37] however, doping itinerant charge carriers into these large gap insulators is a challenge. In the pursuit of an alternative platform that can host a similar triangular lattice network of $Ln$ moments with a smaller, dopable semiconductor charge gap, the $Ln$(Cd,Zn)$_3$(P,As)$_3$ family of compounds presents a number of interesting candidates.[38–56]

*Ln*(Cd,Zn)$_3$(P,As)$_3$ compounds, based on the prototypical ScAl$_3$C$_3$ structure type, host an ideal triangular lattice of rare-earth moments whose anisotropies and ground states can be tuned via *Ln* composition. There is also evidence that they are tunable semiconductors that can be driven from insulators into metals via steric effects[40] or via slight self-doping effects during crystal growth.[42] In contrast to the delafossite-variant *ALnX*$_2$ compounds, strong covalency between the transition metal and pnictide sites leads to a highly disperse valence band, likely enabling *p*-type doping.[39,47,56] An important open question is whether the carriers in these compounds, derived mainly from the transition metal-pnictide interstitial layers, experience comparable frustration effects and whether additional charge or bond order instabilities can be interfaced with the frustrated *Ln* network.

In this paper, we demonstrate that the *Ln*Cd$_3$P$_3$ family of compounds hosts a highly frustrated lattice instability via the formation of local bond order across a honeycomb network rooted in distorted trigonal-planar CdP$_3$ layers within the unit cell. A combination of single-crystal diffuse X-ray scattering and powder pair distribution function (PDF) analyses identify a local shortening of 2/3 of the trigonal Cd–P bonds, resulting in the formation of one-dimensional CdP chains which are prevented from forming long-range order via geometric frustration in a manner analogous to kagome ice correlations.[57] Interactions between these distortions can be mapped to a frustrated tiling of dimers across a honeycomb network,[58,59] leading to fluctuating orthorhombic domains which locally break the in-plane rotational and translational symmetry. Our results establish the *Ln*Cd$_3$P$_3$ family of triangular lattice antiferromagnets as a rare material platform where frustration of the charge degree of freedom can be interfaced with frustrated magnetism within the same lattice framework.

## II.   RESULTS

### A. Local structure and powder PDF

A representative unit cell for $LnCd_3P_3$ is shown in Figure 1(a). The structure is composed of alternating layers of $LnP_6$ octahedra, $CdP_4$ tetrahedra, and $CdP_3$ trigonal planar units [Fig. 1(b)]. The latter form a honeycomb pattern, with Cd and P alternating at the vertices, reminiscent of hexagonal BN and of the Zn(As/P) layer in hexagonal Zn(As/P)[60,61] [Figure 1(c,d)]. Here we denote the two separate Cd crystallographic sites as $Cd_{trig}$ and $Cd_{tet}$; the P sites are denoted in a similar fashion.

In first examining the structure across the $LnCd_3P_3$ compound series, refinement of average structural models using synchrotron powder X-ray diffraction measurements reveal large anisotropic displacement parameters for the $Cd_{trig}$ and $P_{trig}$ sites in all four $Ln$ = La, Ce, Pr, and Nd compounds. Data and refinements are shown in Figure 2(a-d). The values for the structural refinements are reported in the supplementary information (Table S1) and are consistent with a previous report of $PrCd_3P_3$.[38] In all four compounds, $Cd_{trig}$ cations exhibit large displacements along the $c$ axis at room temperature, displacing toward the large voids above and below the site. $P_{trig}$ anions, on the other hand, have larger in-plane displacements at this temperature. Notably, upon cooling toward $T = 5$ K, both the $Cd_{trig}$ cation and $P_{trig}$ anion displacements evolve to be strongly in-plane, revealing a tendency toward two-dimensional disorder within the $ab$ plane.

To explore this further, X-ray PDF measurements of the local structure of polycrystalline samples of $LnCd_3P_3$ were performed at $T = 80$ K [Fig. 2(e-i), Fig. S1] and 300 K (Fig. S2). Small-box modeling and fitting to the data reveal a preference for a distorted, orthorhombic $Cmcm$ cell over the average hexagonal $P6_3/mmc$ cell that is enhanced as the fitting range becomes more local [Fig. 2(g)]. While this $Cmcm$ model allows for in-plane distortions of both the tetrahedral and trigonal planar Cd and P sites, a model including only distortions of the $Cd_{tet}$ and $P_{tet}$ sites shows poorer agreement with the experimental powder PDF compared to a model including only $Cd_{trig}$ and $P_{trig}$ distortions (Fig. S3). This is consistent with the refined in-plane anisotropic displacement parameters at $T = 5$ K, which are strongly suppressed relative to those of $Cd_{trig}$ sites, which remain an order of magnitude larger at these temperatures. Together, these observations suggest that the primary driver of the proposed distortion and symmetry lowering is the breaking of rotational symmetry in the trigonal planar CdP$_3$ units within the CdP layer. This in-plane distortion mode results in a local contraction of two of the three equivalent Cd–P bond lengths within the Cd$_{trig}$ layers and a potential local bond order that is frustrated within the plane [Fig. 2(h,i)].

### B. Local bond order and single-crystal ΔPDF

To further study the presence of local symmetry lowering, single crystal synchrotron X-ray diffraction measurements were performed on crystals of LaCd$_3$P$_3$, CeCd$_3$P$_3$, PrCd$_3$P$_3$, and NdCd$_3$P$_3$. Because the data are qualitatively similar in reciprocal-space structure amongst all four compounds, a representative ($H$, $K$)-plane diffraction pattern for PrCd$_3$P$_3$ is presented in Figure 3(a). Corresponding data for all four compounds are shown in Fig. S4. Note that weak reflections observed at 1/3-type positions are not intrinsic to the sample and are instead a known λ/3 artifact of the X-ray source.[62] Aside from sharp Bragg peaks at integer values of $H$ and $K$, extensive diffuse scattering appears connecting Bragg reflections. The scattering is quasi-two-dimensional and completely diffuse along $L$, generating structured planes of scattering perpendicular to (1, 1, 0) and equivalent reciprocal lattice directions.

Because the diffuse scattering is clearest in the PrCd$_3$P$_3$ dataset, this member was chosen for analysis as representative of the behavior in all four compounds. Line-cuts of the diffuse scattering are presented along directions perpendicular [Fig. 3(d)] and parallel [Fig. 3(e)] to the planes of diffuse scattering for PrCd$_3$P$_3$. Additional line-cuts for the other $Ln$Cd$_3$P$_3$ are presented in Fig. S6. The line-cuts within the plane show that the diffuse plane is structured and peaked at half-integer positions, indicating a local breaking of translational symmetry. We note that the intense features in Figure 3(b) at integer $H$ and $K$ in the half-integer $L$ planes are due to the additive bleed-through from the tails of the primary Bragg peaks at integer $L$ values. These peaks have a diffuse component along $L$, as is often seen in layered materials (See Figures S5 and S8).

The peak shapes were fit using a pseudo-Voigt profile, and the in-plane correlation lengths, extracted via the inverse of the full width at half maximum (FWHM) of the Lorentzian component of the profile, are shown in Figure 3(g). Correlation lengths as well as the intensity of the local correlations are relatively temperature-independent and are seemingly decoupled from thermally induced atomic displacements, suggesting that the diffuse intensity arises from frustrated charge correlations within the trigonal planar CdP layer.

To further resolve the origin of the short-range charge correlations, three dimensional difference pair distribution function (3D-ΔPDF) analysis of the X-ray scattering data[63,64] was performed using a modified approach to the punch-and-fill method.[65] By assessing the real-space features in the 3D-ΔPDF, it is possible to constrain the behavior and identity of the sites participating in the generation of the diffuse scattering in reciprocal space. In particular, one can resolve a real-space map of the vector probabilities of deviations in the charge density relative to the average structure. This can then be modeled using Monte Carlo techniques and compared with the experimentally observed pattern of diffuse scattering.

A representative ($H$, $K$)-plane of the isolated diffuse scattering with the average structure removed and the real part of the Fourier transform are shown for PrCd$_3$P$_3$ at $T$ = 300 K in Figure 4(b,c). The vector probabilities of in-plane correlations in the experimentally-derived 3D-ΔPDF are highest in intensity at integer lattice units in real space, corresponding to distances between like atoms in different unit cells. A quadrupole-like displacement feature is revealed at the first-nearest neighbor ⟨1, 0, 0⟩ positions and another in-phase displacement correlation appears at the ⟨2, 0, 0⟩ positions, indicative of strongly one-dimensional correlations along this direction. Importantly, the orientation of these quadrupoles indicates a change in the interatomic vector component which is orthogonal to the direction connecting these pairs of atomic sites. These correlations decay exponentially and have correlation lengths on the order of 5 unit cells, as demonstrated in Fig. S9. We also observe weaker quadrupole-like features at positions away from the ⟨1, 0, 0⟩-type directions, hinting at displacive correlations along the in-plane direction perpendicular to the direction of the aforementioned one-dimensional correlations.

While the three Cd–P bond lengths in the CdP$_3$ layers are equal in the hexagonal cell, powder PDF results indicate one bond is longer than the other two in the *Cmcm* cell. For instance, in the $T$ = 80 K PrCd$_3$P$_3$ data, the shorter bonds are 2.42(11) Å, whereas the longer bond is 2.53(2) Å, representing an approximate 2% decrease/increase in the bond length with respect to the hexagonal cell. This distortion effectively reduces the CdP$_3$ trigonal planar units to one-dimensional CdP chains and maps the local distortion to a two-dimensional tiling of the honeycomb Cd–P network with "dimers", where each dimer corresponds to the single long-bond within each unit. This type of bond order is expected to be frustrated across the honeycomb network and is a likely candidate for the local in-plane correlations isolated in the 3D-ΔPDF analysis.

To assess the microscopic nature of these correlations, we employed forward Monte Carlo (FMC) modeling to generate a large-box model with short-range correlations between sites on the CdP honeycomb lattice, the expected diffuse scattering, and resulting ΔPDF maps. The dimer covering of the honeycomb network is canonically defined such that each vertex of the honeycomb is reached by exactly one dimer. This problem can be mapped to a 2D Ising model on the (dual) triangular lattice.[58,59] In this mapping, Ising spins decorate the center of each hexagon, and dimers are placed on those edges which sit between frustrated pairs of spins (up-up or down-down) as illustrated in Figure 4 (a-c).

Within the dual model of Ising spins on an antiferromagnetic triangular lattice with couplings up to third nearest neighbor ($J_1$, $J_2$, $J_3$ < 0), in the limit of large nearest neighbor $J_1$, frustration effects manifest one-dimensional "stripe" textures for $J_3/J_2$ < 0.5 and "zig-zag" textures for $J_3/J_2$ > 0.5.[66] The stripe pattern in the Ising model maps to the "staggered" configuration in the dimer model, and the zig-zag pattern in the Ising model maps to a "Herringbone"-like configuration of dimers on the honeycomb lattice [Fig. 1(e-g)].[67,68] The forward Monte Carlo model was built around the 2D Ising problem, and, after relaxing the spin configuration, the coordinates of each Cd$_{trig}$ and P$_{trig}$ site were modified to elongate one of the three Cd–P bonds for each CdP$_3$ unit. The magnitude of these displacements were parameterized based on the orthorhombic $Cmcm$ solution from the aforementioned small box modeling of the powder PDF data, though the model is not constrained by this symmetry. We note that an alternative parameterization was explored via a three-state Potts model, where each Cd$_{trig}$ is assigned a distortion away from one of the three $P_{trig}$ sites, and this model is capable of capturing some of the same patterns of local order as demonstrated in the Supplementary Information (Fig. S11); however, the agreement with the scattering data and ΔPDF is poorer than the Ising model parameterization. There are qualitative differences in the patterns of diffuse scattering expected in the (*H*,*K*)-plane as shown in Figure S11.

By fixing a relatively large $J_1$ and sweeping the ratio of $J_3/J_2$ in the forward Monte Carlo simulation, we are able to tune the short-range correlations in a 100×100×1 supercell between the staggered and Herringbone configurations of the dimer model (Fig. S10). Examination of the calculated diffuse scattering for each case reveals that the diffuse scattering planes observed in the experiment are closely replicated for $J_3/J_2 \geq 0.5$, and the experimentally observed half-integer maxima are recreated for $J_3/J_2 > 0.5$ due to the nature of the symmetry lowering in the Herringbone configuration of dimers.

Optimized Monte Carlo scattering results are presented in Figure 4, where $J_3/J_2$ = 0.58. The resulting dimer model configuration shown in Figure 4(e) features an arrangement of one-dimensional CdP chains which are correlated over short length scales (≈ 3 to 5 unit cells), with a weak tendency toward breaking translational symmetry by alternating chain direction. Due to the three-fold degeneracy of distortion pathways, three domains form, each rotated by 120° relative to the other two. The calculated diffuse scattering that results from this short-range ordered supercell is shown in Figure 4(f). The planes of diffuse scattering, including the modulations in the structure factor and the weak half-integer maxima, are well-captured by the model. For further comparison, the 2D-ΔPDF transformation of this simulated scattering also agrees well with the corresponding 3D-ΔPDF transformation of the experimental scattering data [Fig. 4(g)], replicating the key features at integer coordinates in the experimental 3D-ΔPDF. It is worth noting that including $P_{trig}$ displacements is necessary to replicate the experimentally observed modulation of the structure factor and the corresponding ΔPDF feature at the $Cd_{trig}$-$P_{trig}$ interatomic vectors ⟨-1/2,3/2,0⟩, confirming that both $Cd_{trig}$ and $P_{trig}$ displacements play a role in generating the short-range order.

In order to further strengthen the interpretation presented in our model, reverse Monte Carlo modeling (RMC) was used to refine a large-box structural model against the experimental diffuse scattering. Because the scattering is completely diffuse along $L$, data from the ($H$, $K$, 5/2) scattering plane was used as the target for the RMC model. A 75×75×1 supercell including only a trigonal CdP plane was initialized with a random distribution of Cd and P distortions away from Cd–P bond directions A, B, and C [as defined in Fig. 1(d)], and each RMC move involved a swap of the displacement type of two randomly selected sites of the same element. The termination criteria was defined as 10 times the number of atoms in the supercell. The refined supercell, diffuse scattering, and ΔPDF after convergence are shown in Figure 4(h-j). The RMC generally agrees with the results of the FMC model, replicating a similar chain correlation length and exhibiting a weak tendency to form chains which alternate direction. Taken together, the FMC and RMC results confirm that the origin of the diffuse scattering is bond-dependent, short-range interactions between Cd–P displacive distortions which follow the energy landscape enforced by frustrated bond order and result in a lower symmetry at the local scale. Coupled to this primary, in-plane distortion are weaker, secondary distortions in the tetrahedral layers above and below the Cd$_{trig}$ layers (as shown in supplemental Fig. S7). These do not impact the mapping to the 2D Ising model discussed above.

## III.   DISCUSSION

In $Ln$Cd$_3$P$_3$ compounds, the valence band maximum is composed primarily of phosphorus $p$ states, specifically from those P anions within the CdP layer.[39] Given that the Cd$_{trig}$ cations are bonded solely to the three P anions within the CdP layer, the bonding interactions are, therefore, probably between the $s$ states of Cd$_{trig}$ and $p_x$ and $p_y$ atomic orbitals of P. These atomic orbitals probably hybridize to enable a trigonal planar coordination environment.

In the ionic limit, any degree of charge transfer from the P$^{3-}$ anions, which possess filled $p_x$ and $p_y$ bands (valence band maximum), to the trigonal planar Cd$^{2+}$ cations, which possess empty Cd$^{2+}$ states (conduction band minimum), will result in a second-order Jahn-Teller instability in the P $p$ states. Similarly, in the covalent limit, strong hybridization between the Cd $s$ and P $p$ states would form bonding orbitals which would contain two holes and be subject to a first-order Jahn-Teller instability. In either scenario, a bonding instability is expected to arise from the unconventional trigonal planar CdP$_3$ bonding environment.

The resulting bond instability generates local CdP chains within the honeycomb network whose configurations are governed via chemical bonding considerations – avoiding over- and under-bonded P ions. The effective interactions of this local bond order on a honeycomb lattice frustrate the formation of long-range order and map to the dual triangular lattice with Ising-like interactions.[58,59] An alternative way to envision things is to consider local 120° displacements of Cd ions toward P ions in the honeycomb network. These displacement modes have a local Ising-like degree of freedom (toward or away) and can be mapped onto the midpoints of the CdP honeycomb bond network. The Ising degree of freedom within the honeycomb bond network maps into a kagome network, which is a line graph of the honeycomb lattice. An Ising degree of freedom on this emergent kagome lattice corresponds to the kagome ice problem of antiferromagnetic Ising spins as shown in Figure 4(d). The solutions of this model in the classical limit are equivalent to those attained via dimer tiling on a honeycomb network.[57]

Within the dual triangular lattice Ising model, extended interactions within a $J_1$−$J_2$−$J_3$ model can generate a range of local patterns of order.[66] The "zig-zag" state accessed in the limit of appreciable antiferromagnetic $J_3$ in the triangular lattice corresponds to the Herringbone local order of the dimer model revealed in reverse Monte Carlo analysis of the diffuse scattering in $Ln$Cd$_3$P$_3$. The resulting steric frustration is therefore similar to the frustrated charge order reported in ScV$_6$Sn$_6$[69,70] and related kagome compounds[71–73] with an out-of-plane lattice chain instability. An unusual off-centering of the Cd$_{trig}$ and As$_{trig}$ ions was reported for similar $Ln$Cd$_3$As$_3$ compounds, suggesting a generic in-plane instability for this entire class of materials and warrants further structural investigations.[56]

Preliminary investigations suggest that the frustrated bond instability intrinsic to the nonmagnetic CdP layers indeed couples to the triangular lattice magnetic layers in these compounds. Inelastic neutron scattering data probing the local wave function of the lanthanide ion show that it experiences a local symmetry breaking due to the trigonal planar CdP network's distortions as shown in supplemental Figure S12. This explicitly demonstrates a coupling between the rare earth moments and the local, structural symmetry lowering mediated via the crystalline electric field (CEF) about the rare earth sites. We note here that this coupling may be partially or fully screened in metallic samples, and future measurements are required to ascertain the impact of carrier doping.

The extended interactions mediated via a strain field therefore imply a potentially rich set of states accessible via external strain in the $Ln$Cd$_3$P$_3$ family of compounds. Pronounced strain effects and coupling of lattice fluctuations are known to couple to the charge order state and electronic properties of kagome materials with a similarly frustrated charge ordered state.[74–77] The narrow band gap of the $Ln$Cd$_3$P$_3$ family suggests these compounds are dopable into a metallic state as well. Previous studies of $Ln$Cd$_3$P$_3$ crystals grown via self-flux techniques indeed reported a weakly metallic state with a small carrier density.[42] Crystals grown via this and other methods may be self-doped (hole-doped) via excess phosphorus within the CdP blocks.[42,78] To demonstrate the possibility of hole-doping, a series of intentionally, chemically doped Pr$_{1-x}$Sr$_x$Cd$_3$P$_3$ crystals were created and the transition into a metallic state is shown in supplemental Figure S13.

More generally, the ability to dope the $Ln$Cd$_3$P$_3$ compounds opens a fascinating new frontier of exploring carrier tuning in a lattice hosting both strong magnetic and steric frustration. A quantum disordered magnetic ground state hosted within the local moments of the $S_{eff}$ = 1/2 $Ln$ layer can, in principle, be interfaced with the frustrated bond order within the CdP$_3$ planes via charge carriers, potentially leading to a number of interesting effects. These could include giant magnetocaloric responses[79,80] in the magnetic channel or emergent behavior, such as unconventional superconductivity[81,82] within the charge channel. Future experiments on electronic symmetry breaking in carrier-doped $Ln$Cd$_3$P$_3$ are therefore highly desired.

In conclusion, we have studied the $Ln$Cd$_3$P$_3$ family of compounds and have identified a frustrated form of bond order akin to kagome ice within its honeycomb, trigonal planar CdP layer. The bond frustration derives from a chain instability within the honeycomb network and therefore maps to frustrated Ising spins on a dual triangular lattice model. Signatures of this frustration persist to room temperature and exist independently from the magnetic frustration within the triangular lattice of $Ln$ $S_{eff}$ = 1/2 moments. Our findings motivate future experiments probing the coupling between the two (bond and moment) frustrated networks as well as the impact of carrier doping as a means of stabilizing emergent phase behavior in this fascinating family of compounds.

## IV. ACKNOWLEDGMENTS

SDW acknowledges helpful discussions with Leon Balents and Jacob Ruff. The authors acknowledge various forms of support from Guang Wu, Matt Krogstad, Joe Paddison, Jose Marquez, and Casandra Gomez Alvarado. This work was supported by the US Department of Energy (DOE), Office of Basic Energy Sciences, Division of Materials Sciences and Engineering under Grant No. DE-SC0017752. S.J.G.A. acknowledges additional financial support from the National Science Foundation Graduate Research Fellowship under Grant No. 1650114. J.R.C. acknowledges support through the NSF MPS-Ascend Postdoctoral Fellowship (DMR-2137580). This research made use of the shared facilities of the NSF Materials Research Science and Engineering Center at UC Santa Barbara (DMR-2308708). Use was made of computational facilities purchased with funds from the National Science Foundation (CNS-1725797) and administered by the Center for Scientific Computing (CSC). The CSC is supported by the California NanoSystems Institute and the Materials Research Science and Engineering Center (MRSEC; NSF DMR-2308708) at UC Santa Barbara. G.P., B.R.O., and L.K. acknowledge support from the National Science Foundation (NSF) through Enabling Quantum Leap: Convergent Accelerated Discovery Foundries for Quantum Materials Science, Engineering and Information (Q-AMASE-i): Quantum Foundry at UC Santa Barbara (DMR-1906325). J.H. acknowledges financial support from the Bavarian Californian Technology Center (BaCaTeC). This research used resources of the Advanced Photon Source, a U.S. Department of Energy (DOE) Office of Science user facility operated for the DOE Office of Science by Argonne National Laboratory under Contract No. DE-AC02-06CH11357. Research conducted at the Center for High-Energy X-ray Sciences (CHEXS) is supported by the National Science Foundation (BIO, ENG and MPS Directorates) under award DMR-2342336.

## V. AUTHOR CONTRIBUTIONS

J.R.C., A.R.J., D.R., C.B., and J.H. contributed to sample synthesis. J.R.C. and L.C.G. performed synchrotron powder X-ray diffraction data collection. S.J.G.A, J.R.C., R.S, and S.D.W. contributed to PDF analysis. S.J.G.A, G.P., R.G., B.R.O., S.Sa., and L.K. contributed to single crystal X-ray diffraction data collection and processing, and S.S., S.D.W., and D.R. performed analysis of the neutron scattering data. S.J.G.A. performed single crystal diffuse X-ray scattering analysis and $\Delta$PDF analysis. D.R. and J.H. performed bulk property measurements, D.R., M.B.S, and V.O.G. performed neutron scattering measurements. S.J.G.A., J.R.C, and S.D.W. prepared the manuscript and analyzed data with input from all other co-authors.

## VI. COMPETING INTERESTS

The authors declare no competing interests.

## VII. FIGURES

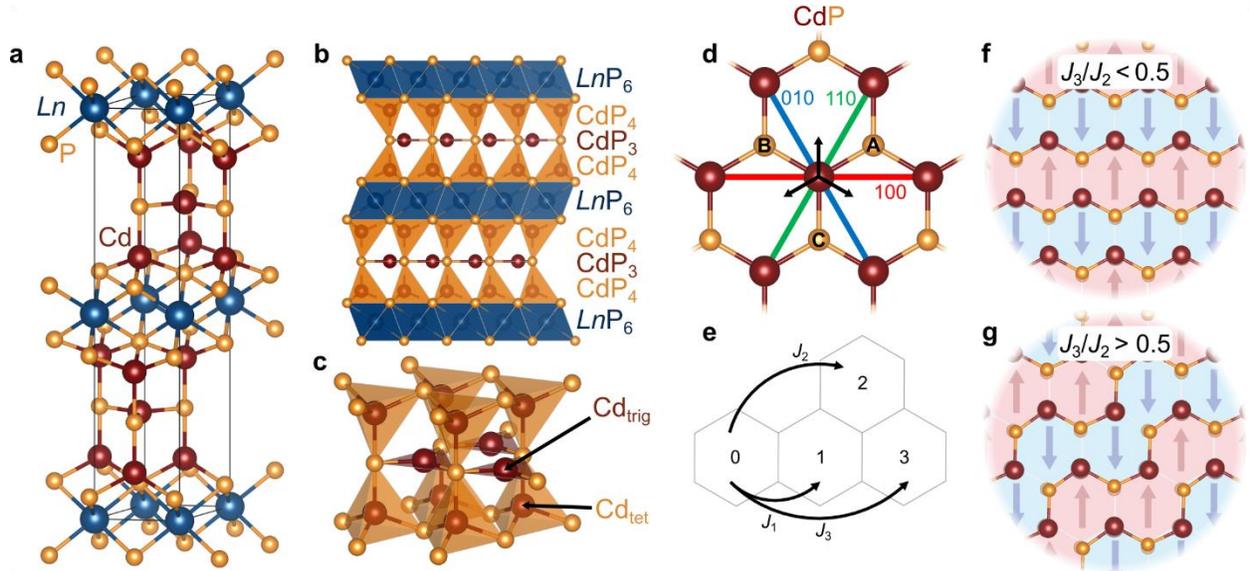

**Fig. 1 | Structure and bond instability in *Ln*Cd$_3$P$_3$. a** The unit cell of the ScAl$_3$C$_3$-type structure of *Ln*Cd$_3$P$_3$ (space group *P*6$_3$/*mmc*). **b** The structure is composed of alternating layers of *Ln*P$_6$ octahedra, CdP$_4$ tetrahedra, and CdP$_3$ trigonal planar units. **c** We distinguish the trigonal planar and tetrahedral Cd sites as Cd$_{trig}$ and Cd$_{tet}$, respectively. **d** The distortion modes observed in the local structure can be understood as the net displacement of the Cd$_{trig}$ site away from one of three equivalent Cd–P bond directions (labeled A, B, and C), resulting in two short bonds and one long bond. **e** Interactions between distortions on neighboring sites can be mapped to a frustrated dimer tiling across the edges of a honeycomb network, which can, in turn, be mapped to a 2D Ising model on a dual triangular lattice of hexagons (See Fig. S12 for more details). The dimer maps to the single long bond within each CdP$_3$ unit. **f** In the limit $J_1 \rightarrow -\infty$, the dimers (spins) form a staggered (stripe) pattern for $J_3/J_2 < 0.5$, and **g** a Herringbone (zig-zag) pattern for $J_3/J_2 > 0.5$. Representative displacement amplitudes and directions are shown for both Cd$_{trig}$ and P$_{trig}$ sites.

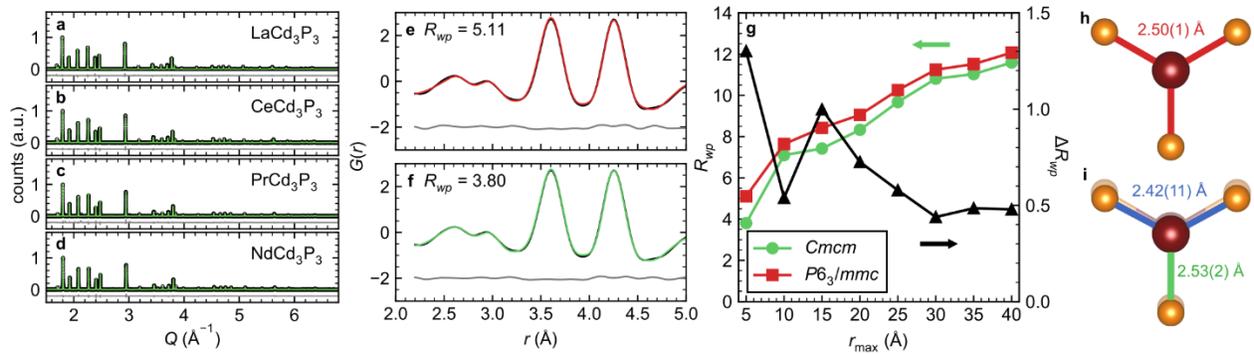

**Fig. 2 | Synchrotron powder X-ray diffraction and pair distribution function analysis. a** Data sets collected at $T$ = 5 K and hexagonal $P6_3/mmc$ fits from Rietveld refinement for LaCd$_3$P$_3$, **b** CeCd$_3$P$_3$, **c** PrCd$_3$P$_3$, and **d** NdCd$_3$P$_3$. Open circles represent raw data, and solid lines indicate fit and difference profiles. **e** Profile fit obtained from small-box modeling for the range 2.2 Å < $r$ < 5 Å, using the hexagonal $P6_3/mmc$ cell and **f** the orthorhombic $Cmcm$ cell. **g** Comparison of the weighted profile $R$-factor ($R_{wp}$) values obtained over the ranges 2.2 Å < $r$ < $r_{max}$ as a function of $r_{max}$. The right axis shows the difference between the $Cmcm$ and $P6_3/mmc$ fits for each $r_{max}$. **h** The refined bond lengths obtained from the $P6_3/mmc$ and **i** $Cmcm$ fits.

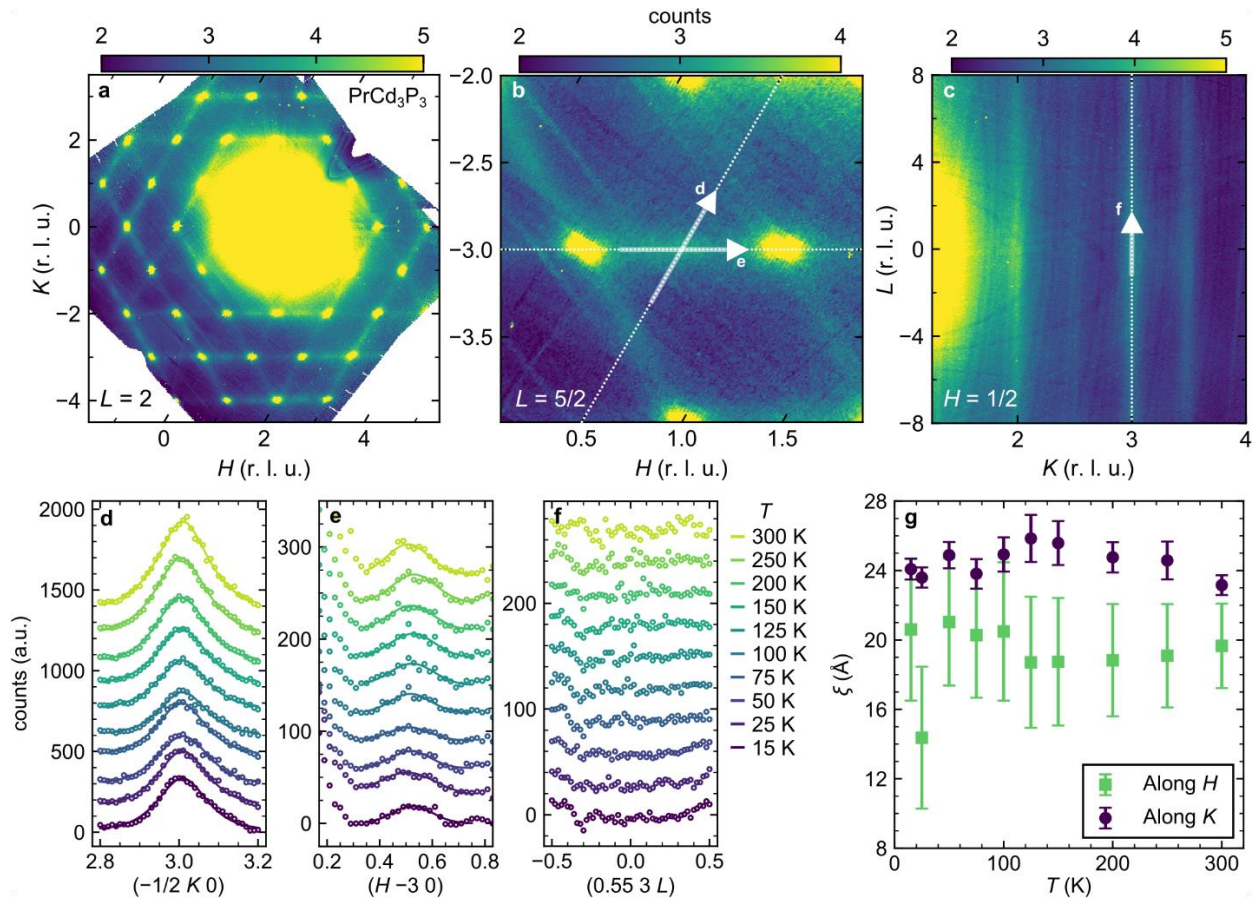

**Fig. 3 | Single crystal diffuse X-ray scattering data and analysis. a** Data collected at $T$ = 250 K in the ($H$, $K$, 2) scattering plane for PrCd$_3$P$_3$ revealing a complex diffuse scattering pattern between Bragg reflections that is representative of the $Ln$Cd$_3$P$_3$ ($Ln$ = La, Ce, Pr, Nd) family. $L$ = 2 was selected for low background intensity. The large diffuse spot near the origin of the dataset is due to small-angle background scattering. **b** Line-cut directions highlighted in a representative ($H$, $K$) scattering plane and **c** ($K$, $L$) scattering plane at $T$ = 250 K. **d** Peak shapes along the (1/2, $K$, 0) line-cut and **e** ($H$, -3, 0) line-cut, each fit using a pseudo-Voigt profile and a linear background which was subtracted before plotting. **f** Line-cut along $L$ showing little to no modulation of the peak intensity along the interplane direction. A linear background was subtracted before plotting. **g** Correlation lengths obtained from the full width at half maximum of the fit profiles in the ($H$, $K$) plane reveal negligible temperature dependence along both directions. Points represent parameter values obtained from non-linear least-squares fitting, and error bars indicate the 1$\sigma$ uncertainties from the fit.

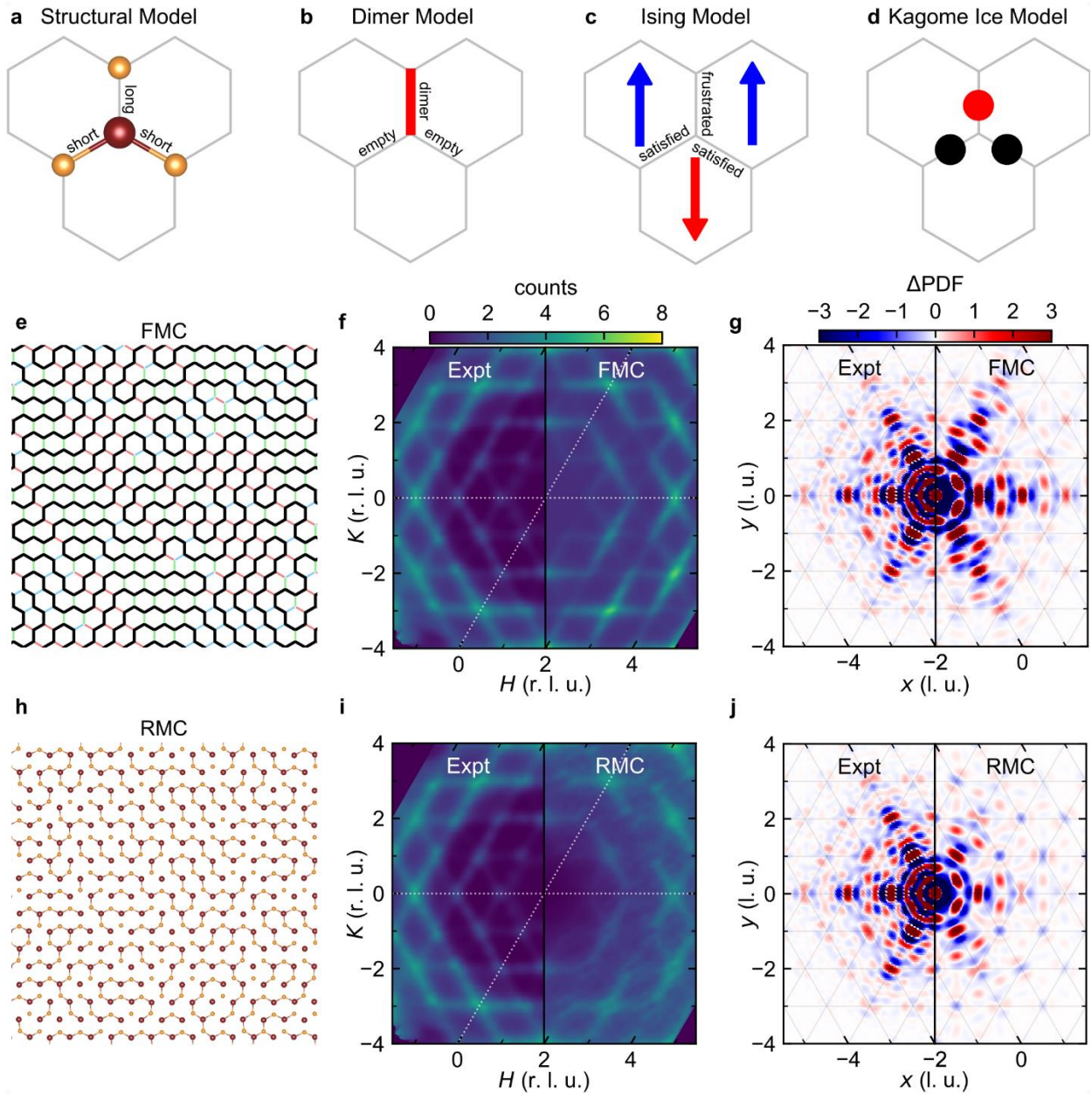

**Fig. 4 | Monte Carlo modeling of short-range bond order, diffuse scattering, and ΔPDF. a–d**, Mapping between the structural model of long Cd–P bonds (**a**) to hard-core dimers on a honeycomb network (**b**), to frustrated bonds in the dual triangular lattice Ising model (**c**), and to a local Ising variable at the midpoint of each bond, generating kagome ice correlations (**d**). **e**, An example of the thermalized ground state resulting from the Forward Monte Carlo (FMC) simulation, revealing short-range, one-dimensional CdP chains. The three degenerate orientations of the long bond are highlighted in different colors, corresponding to the dimers in the forward Monte Carlo model. **f**, Representative

experimental data showing diffuse scattering in the ($H$, $K$, 5/2) scattering plane for PrCd$_3$P$_3$ at $T$ = 300 K isolated by a modified punch-and-fill method, and the calculated diffuse scattering from the thermalized ground state of the FMC model. **g**, The 3D-ΔPDF for PrCd$_3$P$_3$ extracted from the diffuse scattering at $T$ = 300 K, and the 2D-ΔPDF resulting from the FMC model. **h**, An example of the refined supercell configuration of CdP chains after reverse Monte Carlo (RMC) modeling using the ($H$, $K$, 5/2) scattering planes as the target. **i**, Comparison of the experimental diffuse scattering and the calculated diffuse scattering form the refined RMC model. **j**, Comparison of the experimental 3D-ΔPDF and the calculated 2D-ΔPDF obtained from the RMC model.

## IX. METHODS

### A. Sample synthesis

Single crystals of $Ln$Cd$_3$P$_3$ were prepared from a molten salt flux.[38] A stoichiometric amount of elemental $Ln$ powder (Fisher Scientific, 99.9% REO), Cd$_3$P$_2$ powder (Alfa Aesar, 99.5% metals basis), and red P (BeanTown Chemical, 99.9999%) was combined in a total of 500 mg of solute powder. This powder was then sealed in an evacuated silica ampoule with 2 g of 1:1 molar ratio of KCl and NaCl salts. The silica ampoule was then heated at 50 °C/hr to 900 °C, where it dwelled for 72 hours. Then, it was slowly cooled to 660 °C at 2 °C/hr, then allowed to furnace cool to room temperature. To retrieve the samples, the silica ampoules were first scored and opened above the solidified salt. The section of the ampoule with the solidified salt was then placed in a beaker full of deionized water on a hot plate so the salt could be dissolved. As the salt dissolved, the crystals and other products would accumulate at the bottom of the beaker. Most of the excess dissolved salt and water were discarded, and the crystals and products were examined and extracted.

Polycrystalline samples were prepared using conventional solid-state synthetic techniques.[41] First, $Ln$P precursors were prepared by direct reaction of elemental lanthanide powders with red P. Stoichiometric amounts of the elements were weighed and ground together in an Ar-filled glovebox, followed by heating in alumina crucibles at 850 °C for 48 hours in sealed, evacuated silica ampoules. These precursors were then combined with a stoichiometric amount of Cd$_3$P$_2$, and ball milled in a Spex 8000D Mixer/Mill for 1 hour using a tungsten carbide vial and tungsten carbide balls. The milled mixture was then sealed in an evacuated silica ampoule and heated to 800 °C for 24 hours before being quenched in water.

To synthesize powders for neutron scattering measurements isotopically enriched $^{114}$Cd (99%, Trace Sciences Intl. Inc.) was used as a reagent in a modified synthesis process. The isotopic $^{114}$Cd ingot was cut into small pieces inside an Ar-filled glove box and combined with Pr metal filings (Thermo Scientific Chemicals, 99.9 REO, 200 mesh) in a 1:2 Pr:Cd stoichiometric ratio. This mixture was placed in an alumina crucible and sealed in a silica ampoule at an Ar pressure of 256 Torr and reacted at 850 °C for 48 hours before the furnace was switched off to cool to room temperature. Then, this PrCd$_2$ precursor was ground and added to $^{114}$Cd and red P to make the final precursor. This mixture was placed in an alumina crucible, sealed in a silica ampoule at an Ar pressure of 256 Torr, and reacted at 800 °C for 24 hours before quenching in water.

To synthesize hole-doped single crystals of Pr$_{1-x}$Sr$_x$Cd$_3$P$_3$, the aforementioned NaCl/KCl flux growth technique was employed with the exception of the use of a pre-made PrCd$_2$ precursor as the Pr source and a pre-made SrCd precursor as the Sr source. The PrCd$_2$ precursor was synthesized by combining filings from a Pr metal rod (Thermo Scientific Chemicals, 98.5% metals basis, excluding Ta, 6.35 mm DIA) with Cd shot (Alfa Aesar, 99.999% metals basis, 3–6 mm) in a 1:2 Pr:Cd stoichiometric ratio. The SrCd precursor was synthesized by combining stoichiometric amounts of Sr ingot (Alfa Aesar, 99.95% trace metal basis, distilled dendritic pieces) and Cd shot in a tantalum foil which was crimped closed and then loaded inside a steel tube under an Ar pressure of 340 Torr to prevent reaction with the container. The SrCd melt was then reacted at 600 °C for 48 hours and then furnace cooled to room temperature. The PrCd$_2$, SrCd, Cd shot, and red P powder were mixed inside an Ar-filled glove box. A 2% excess by weight of phosphorous was included to compensate for losses due to adhesion to the weighing boat. The mixture was then loaded into an alumina crucible and sealed inside a carbon-coated silica ampoule (3 mm wall thickness) under 256 Torr of Ar. The ampoule was heated to 660 °C at a rate of 50 °C/h and held at that temperature for 24 hours. It was then further heated to 950 °C, dwelled for 72 hours, and subsequently cooled to 660 °C at a rate of 1 °C/h. Final cooling to room temperature was carried out at 200 °C/h. The resulting crystals were extracted by dissolving the salt flux in deionized water. Large, mm-scale crystals were obtained, with thicknesses ranging from 20 to 50 μm.

### B. X-ray scattering measurements

Powder pair distribution function datasets were collected at the 11-ID-B beamline of the Advanced Photon Source (APS), Argonne National Laboratory, with an incident wavelength of 0.1432 Å. Data were collected at $T$ = 80 K and 300 K, and PDF patterns were extracted with a maximum momentum transfer of $|Q|_{max}$ = 25 Å$^{-1}$. Modeling and fitting of the PDF data were performed using TOPAS ACADEMIC[83] and ISODISTORT.[84]

Temperature-dependent single crystal synchrotron X-ray diffraction experiments were performed at the QM$^2$ beamline at the Cornell High Energy Synchrotron Source (CHESS).[62] To avoid the absorption edges of constituent elements, an incident X-ray wavelength λ = 0.676 Å (E = 26 keV) was chosen using a double-bounce diamond monochromator. A stream of cold He gas flowing across the crystal sample was used to control the temperature. A Pilatus 6M area detector array collected the scattering intensities in transmission geometry, and data were collected while the samples were rotated with three tilted 360° rotations, sliced into 0.1° frames. Data were visualized and analyzed using the NEXPY and NXS-ANALYSIS-TOOLS software suites.[85,86] The viewing axes for the (H, K) scattering planes were skewed by 60° to reflect the relative orientation between the H and K axes.

### C. ΔPDF analysis

The single crystal X-ray diffraction data were symmetrized using the symmetry operations belonging to the 6/*mmm* ($D_{6h}$) point group. The full three-dimensional scattering volume used for three-dimensional difference pair distribution function (3D-ΔPDF) analysis was defined by the limits $H_{max}$ = 4.5, $K_{max}$ = 4.5, and $L_{max}$ = 8.5. The large background centered at the origin of reciprocal space was removed from the dataset by subtracting broad, three-dimensional Gaussian profiles. Pixels with counts below a chosen minimum were then filled in with a flat background prior to isolating the diffuse scattering. Bragg peaks and artifacts were removed from the dataset by defining an ellipsoidal volume around each integer (H, K, L) coordinate, and any pixels within this volume possessing a value above a chosen threshold intensity were deleted. Resulting gaps in the data were then filled in via convolution with a Gaussian kernel. Care was taken to ensure that the kernel was circular in the (H, K) scattering plane after a 60° skew of the axes, so that any smoothing effects would impart minimal distortions to the symmetry of the observed scattering. A hexagonal Tukey window was applied to the in-plane scattering with α = 0.5. Along L, a standard Tukey window was applied with α = 1.0. The viewing axes for the (x, y) ΔPDF planes shown in this paper are skewed by 120° to reflect the angle between the crystallographic *a* and *b* axes.

### D. Neutron scattering experiments and crystalline electric field analysis

Neutron scattering measurements were carried out at the Spallation Neutron Source (SNS) at Oak Ridge National Laboratory (ORNL). High energy inelastic scattering data were collected on the SEQUOIA direct-geometry time-of-flight spectrometer. Approximately 1 g of phase-pure, isotopically enriched polycrystalline sample of Pr$^{114}$Cd$_3$P$_3$ was placed in a ¼" inner diameter aluminum cylindrical can and an Al absorption correction was applied to the data. The data were acquired in high-resolution mode using incident energies of $E_i$ = 4, 8, 11, 25, 50, 80, 100, and 200 meV at temperatures of $T$ = 5, 50, 100, and 300 K. All datasets were collected with Fermi Chopper 1 at 120 Hz, and the background contribution from the aluminum sample canister was removed by measuring the same empty canister at each of the above-mentioned incident energies ($E_i$) and temperatures ($T$). Additionally, field-dependent inelastic scattering data were collected using the HYSPEC spectrometer at SNS, ORNL. An incident energy $E_i$ = 9.5 meV and a chopper frequency of 420 Hz was used to probe the crystalline electric field excitations at applied magnetic fields of $\mu_0 H$ = 0, 3, and 5 T and at variable temperatures up to 150 K. The same sample can was loaded into a cryomagnet for these measurements, and empty can background measurements were carried out to subtract the contribution coming from the sample holder.

A point charge model of PrCd$_3$P$_3$ was calculated using the PyCrystalField software[87] based on the average, $P6_3/mmc$ structure with results shown in Table S2 and S3. The resolution function of the SEQUOIA beamline, taking into account the chopper frequency, incident energy, and beamline geometry, was modeled using MCViNE.[88] Line-cuts of data were averaged from $|Q|$ = 1 to 2 Å$^{-1}$ for analysis and fits of the CEF scheme using PyCrystalField. The level scheme determined via fits is shown in Tables S2–S4. The axes chosen for the point charge model and subsequent fits in real space were $x$ = [1, 1/2, 0], $y$ = [0, 1, 0], and $z$ = [0, 0, 1].

### E. Electrical transport measurements

Electrical resistivity and Hall data were collected using a Quantum Design Physical Property Measurement System (PPMS). Crystals were mounted to an AC resistivity puck and contacted using gold wire and silver paint. Measurements were performed using a 4-wire measurement and the electrical transport option (ETO) of the PPMS.

## X. DATA AVAILABILITY

Data underlying the figures and conclusions of this work are publicly available via Zenodo (https://doi.org/10.5281/zenodo.14613498).[89]

# Supplementary Information

I. Average Structural Refinements

Table S1. Structural and refinement parameters for $Ln$Cd$_3$P$_3$ obtained using synchrotron powder X-ray diffraction at $T$ = 300 K and 5 K ($T$ = 5 K data was not collected for CeCd$_3$P$_3$).

|  | LaCd$_3$P$_3$ | | CeCd$_3$P$_3$ | PrCd$_3$P$_3$ | | NdCd$_3$P$_3$ | |
| --- | --- | --- | --- | --- | --- | --- | --- |
| $T$ | 300 K | 5 K | 300 K | 300 K | 5 K | 300 K | 5 K |
| Lattice | | | | | | | |
| $a$ (Å) | 4.294532 | 4.289074 | 4.279790 | 4.268984 | 4.264544 | 4.260452 | 4.256547 |
| $c$ (Å) | 21.088651 | 21.010235 | 20.988965 | 20.947783 | 20.883543 | 20.917044 | 20.844464 |
| | | | | | | | |
| $Ln$ | | | | | | | |
| $U_{11}$ | 0.00757 | 0.00256 | 0.00742 | 0.00614 | 0.00336 | 0.0099 | 0.00461 |
| $U_{12}$ | 0.003785 | 0.00128 | 0.00371 | 0.00307 | 0.00168 | 0.00495 | 0.002305 |
| $U_{33}$ | 0.01028 | 0.01028 | 0.01393 | 0.01833 | 0.01906 | 0.01505 | 0.01415 |
| | | | | | | | |
| Cd$_{tet}$ | | | | | | | |
| $z$ | 0.12814 | 0.128 | 0.12764 | 0.12735 | 0.12714 | 0.12708 | 0.12674 |
| $U_{11}$ | 0.01604 | 0.00414 | 0.01549 | 0.01657 | 0.00686 | 0.01694 | 0.00872 |
| $U_{12}$ | 0.00802 | 0.00207 | 0.007745 | 0.008285 | 0.00343 | 0.00847 | 0.00436 |
| $U_{33}$ | 0.00788 | 0.00687 | 0.01222 | 0.01326 | 0.0125 | 0.01297 | 0.00808 |
| | | | | | | | |
| Cd$_{trig}$ | | | | | | | |
| $U_{11}$ | 0.02475 | 0.03596 | 0.02533 | 0.02826 | 0.02769 | 0.02773 | 0.02448 |
| $U_{12}$ | 0.012375 | 0.01798 | 0.012665 | 0.01413 | 0.013845 | 0.013865 | 0.01224 |
| $U_{33}$ | 0.04703 | 0.01885 | 0.04872 | 0.0501 | 0.01695 | 0.05463 | 0.01439 |
| | | | | | | | |
| P$_{tet}$ | | | | | | | |
| $z$ | 0.5797 | 0.5793 | 0.57869 | 0.57777 | 0.57688 | 0.57739 | 0.57609 |
| $U_{11}$ | 0.00983 | 0.00674 | 0.00868 | 0.00843 | 0.01062 | 0.01299 | 0.01302 |
| $U_{12}$ | 0.004915 | 0.00337 | 0.00434 | 0.004215 | 0.00531 | 0.006495 | 0.00651 |
| $U_{33}$ | 0.02167 | 0.01251 | 0.01616 | 0.01661 | 0.01721 | 0.01431 | 0.01069 |
| | | | | | | | |
| P$_{trig}$ | | | | | | | |
| $U_{11}$ | 0.02346 | 0.03474 | 0.03287 | 0.0392 | 0.03564 | 0.03112 | 0.02137 |
| $U_{12}$ | 0.004915 | 0.00456 | 0.016435 | 0.0196 | 0.01782 | 0.01556 | 0.010685 |
| $U_{33}$ | 0.01001 | 0.02035 | 0.00824 | 0.01844 | 0.01221 | 0.01479 | 0.01614 |

## II. Local Structural Refinements

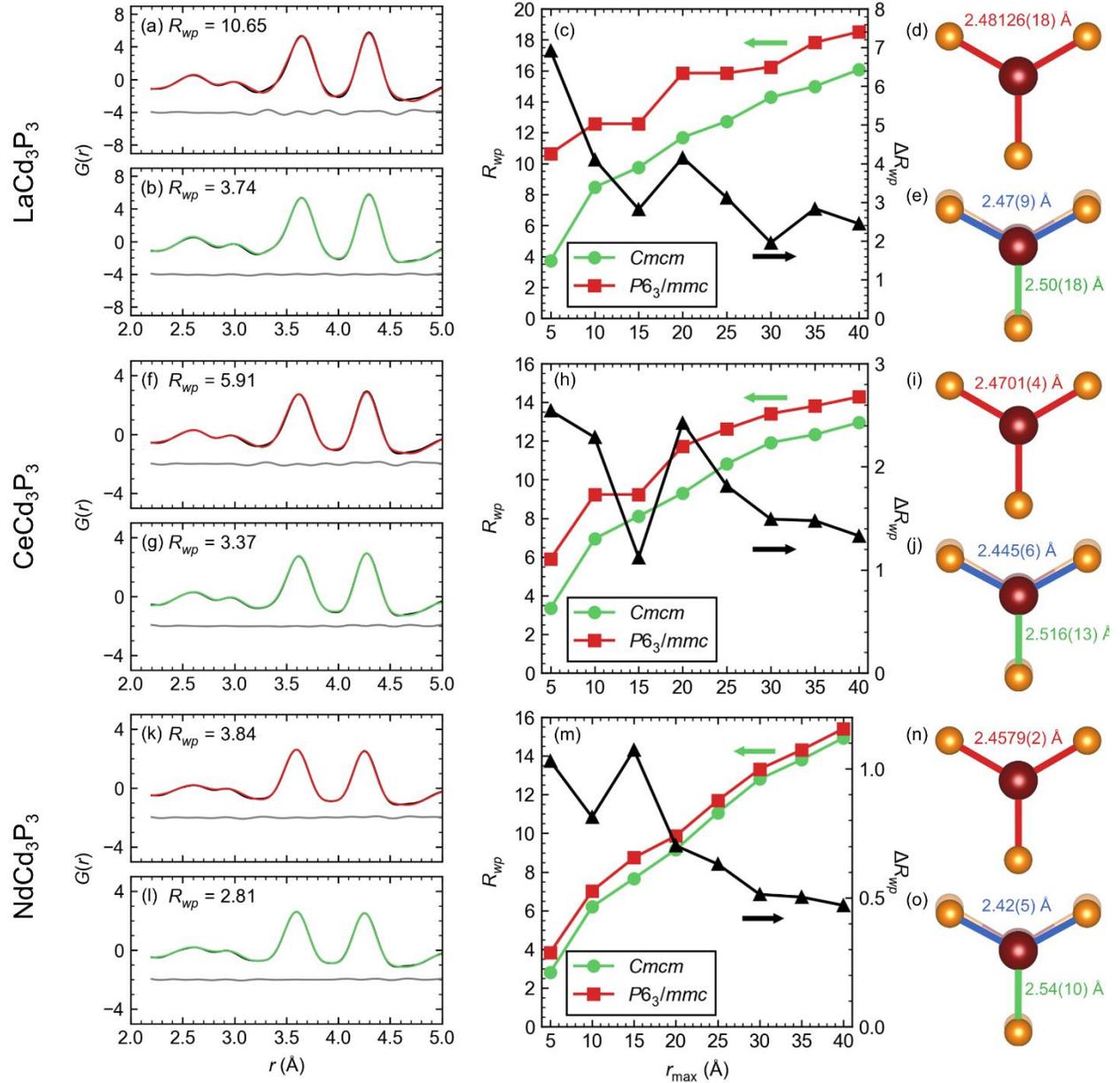

Fig. S1. Powder pair distribution function analysis for LaCd$_3$P$_3$, CeCd$_3$P$_3$, and NdCd$_3$P$_3$ at $T$ = 80 K. (a,f,k) Profile fit obtained from small-box modeling for the range 2.2 Å < $r$ < 5 Å, using the hexagonal $P6_3/mmc$ cell and (b,g,l) the orthorhombic $Cmcm$ cell. (c,h,m) Comparison of the $R_{wp}$ values obtained over the ranges 2.2 Å < $r$ < $r_{max}$ as a function of $r_{max}$. The right axis shows the difference between the $Cmcm$ and $P6_3/mmc$ fits for each $r_{max}$. (d,i,n) The refined bond lengths obtained from the $P6_3/mmc$ and (e,j,o) $Cmcm$ fits.

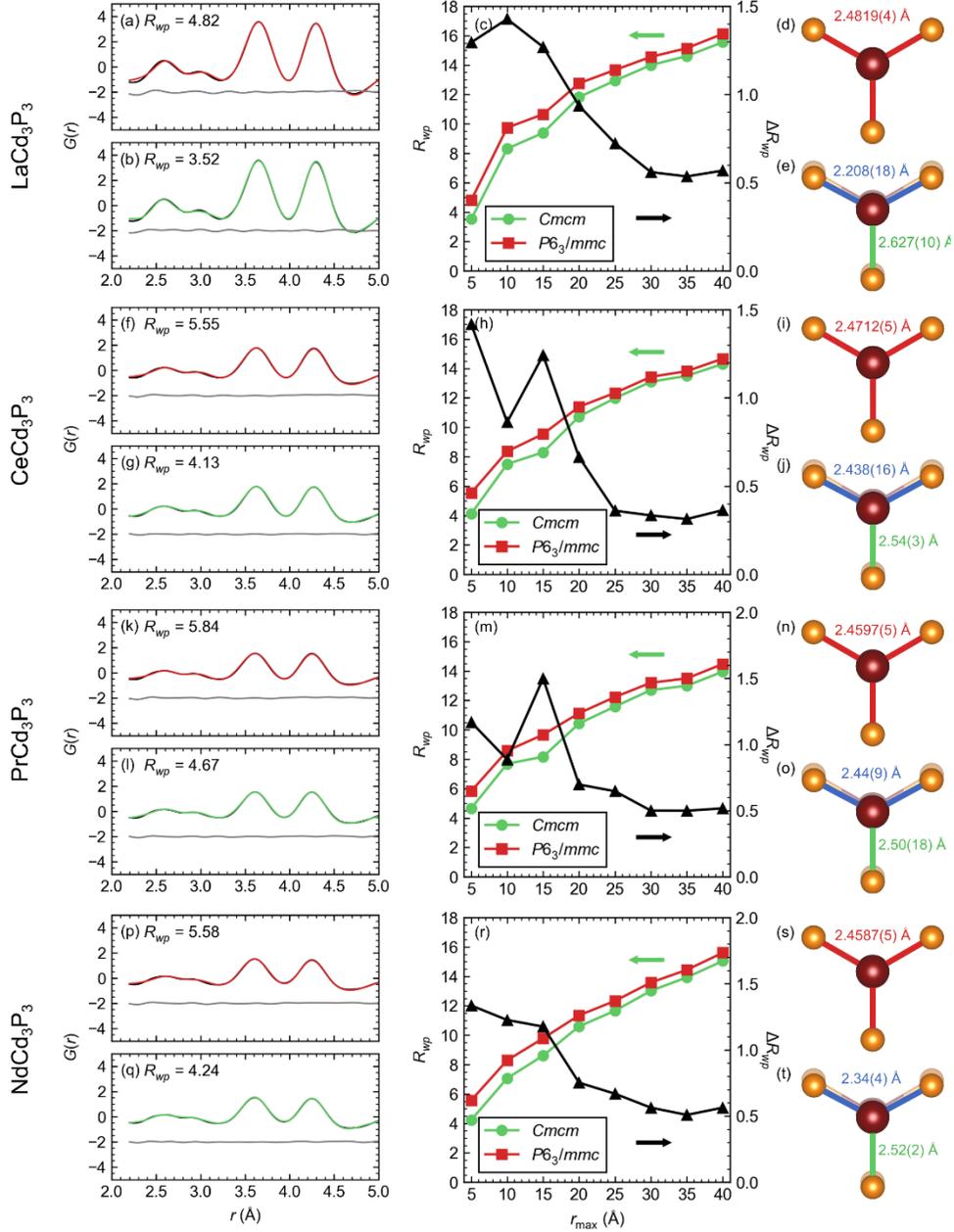

Fig. S2. Powder pair distribution function analysis for LaCd$_3$P$_3$, CeCd$_3$P$_3$, PrCd$_3$P$_3$, and NdCd$_3$P$_3$ at $T$ = 300 K. (a,f,k,p) Profile fit obtained from small-box modeling for the range 2.2 Å < $r$ < 5 Å, using the hexagonal $P6_3/mmc$ cell and (b,g,l,q) the orthorhombic $Cmcm$ cell. (c,h,m,r) Comparison of the $R_{wp}$ values obtained over the ranges 2.2 Å < $r$ < $r_{max}$ as a function of $r_{max}$. The right axis shows the difference between the $Cmcm$ and $P6_3/mmc$ fits for each $r_{max}$. (d,i,n,s) The refined bond lengths obtained from the $P6_3/mmc$ and (e,j,o,t) $Cmcm$ fits.

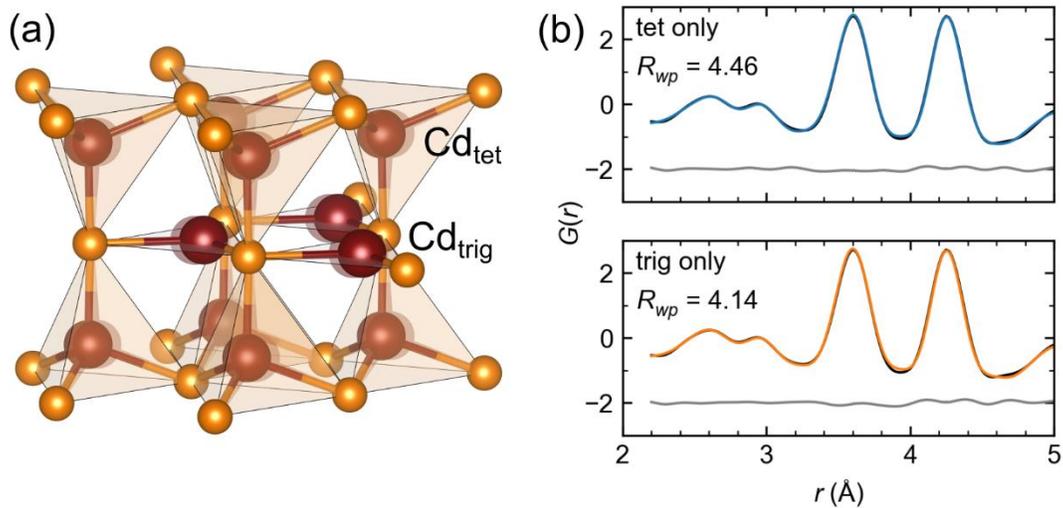

Fig. S3. (a) Schematic of the in-plane distortion modes included in the *Cmcm* model. (b) Comparison of powder pair distribution function models for $PrCd_3P_3$ over the local range 2.2 Å < $r$ < 5 Å including in-plane distortions of the Cd and P sites in the tetrahedral layer only (top) and trigonal layer only (bottom).

## III. Single Crystal X-ray Diffraction Datasets

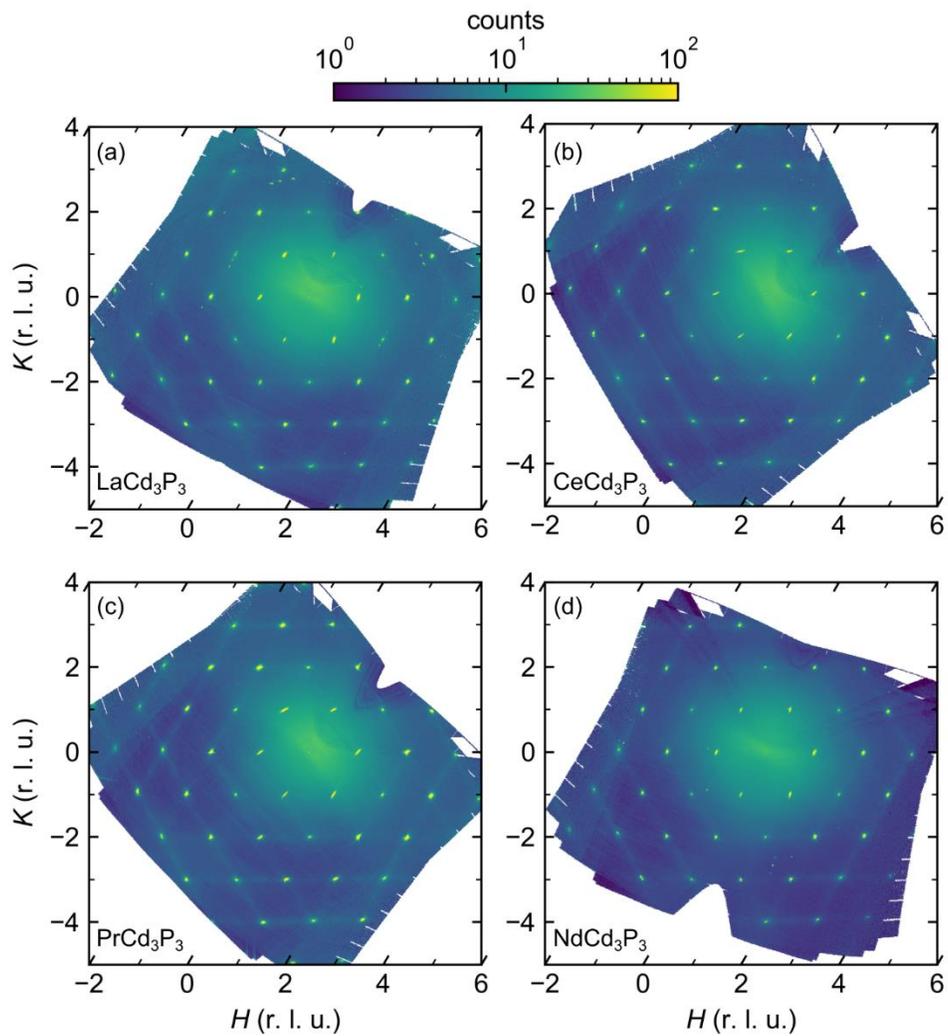

Fig. S4. Single crystal X-ray diffraction in the ($H$, $K$) scattering plane for (a) LaCd$_3$P$_3$, (b) CeCd$_3$P$_3$, (c) PrCd$_3$P$_3$, and (d) NdCd$_3$P$_3$ at $T$ = 300 K.

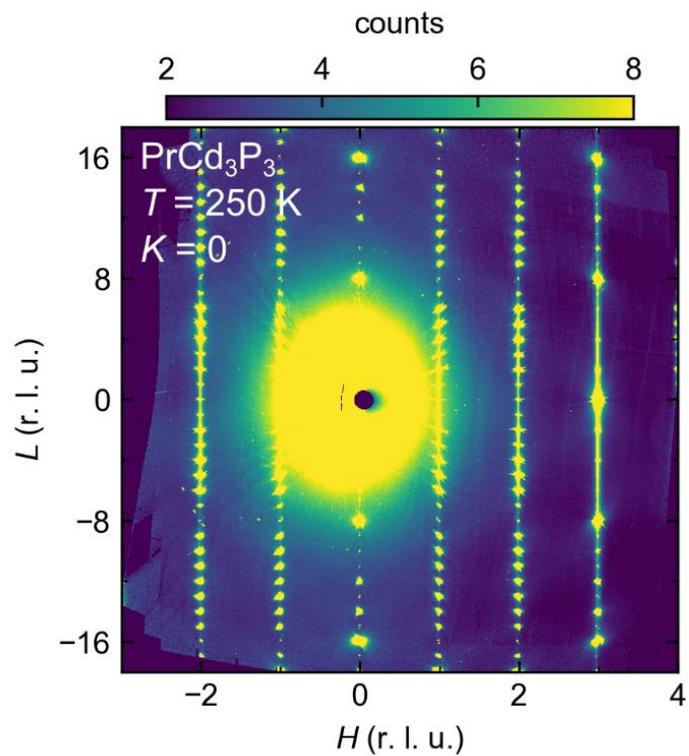

Fig. S5. Single crystal X-ray diffraction in the ($H$, 0, $L$) scattering plane for PrCd$_3$P$_3$ at $T$ = 250 K.

## IV. Diffuse Scattering Analysis

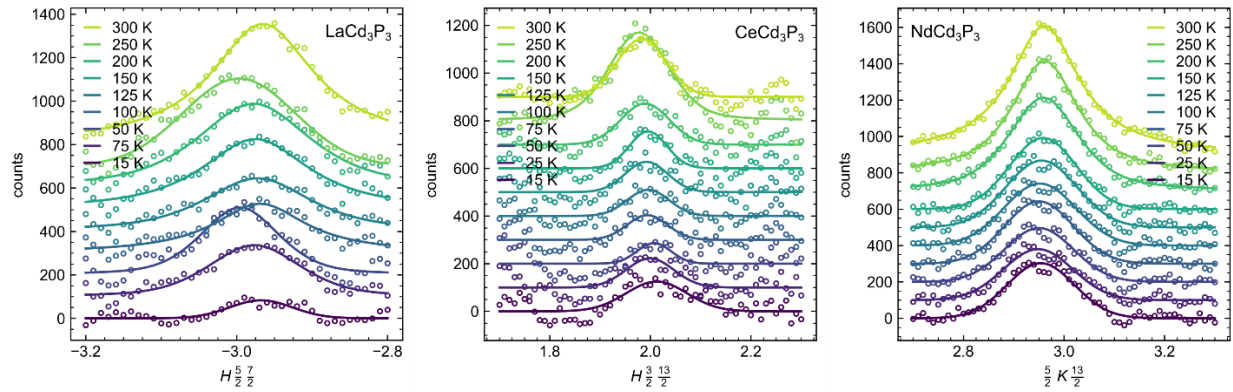

Fig. S6. Line-cuts perpendicular to the planes of diffuse scattering for other $Ln$Cd$_3$P$_3$ compounds.

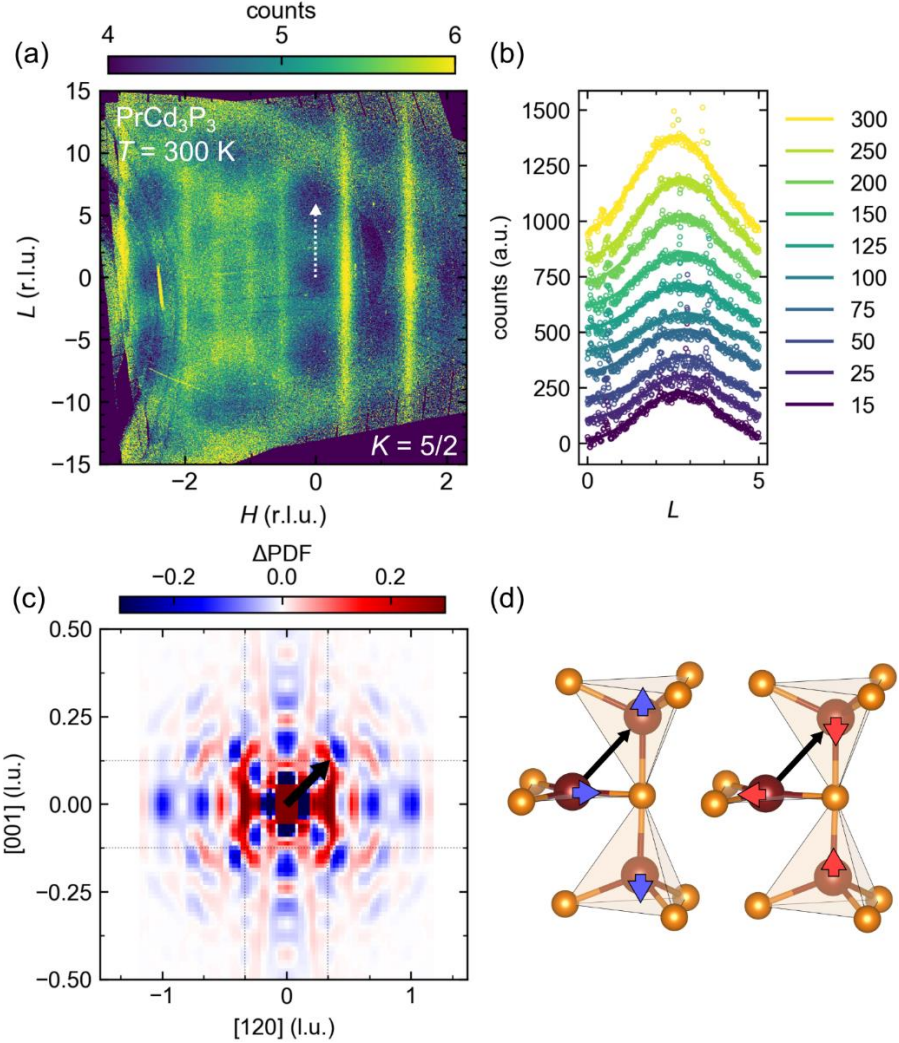

Fig. S7. (a) Diffuse scattering in the ($H$, 5/2, $L$) scattering plane, revealing a broad, quasi-two-dimensional diffuse feature diagonal to the in-plane correlations, hinting at qualitatively weaker inter-layer correlations with a significantly shorter correlation length. (b) Temperature-dependence of the line-cuts of the diagonal diffuse plane along $L$. (c) 3D-ΔPDF in the ($a$, −$a$, $c$) real space plane, revealing the origin of these diffuse planes as a quadrupole-like feature at the ⟨1/3, 2/3, 1/8⟩ positions in real space. Dashed lines mark the ±1/3 and ±1/8 positions along the [1, −1, 0] and [0, 0, 1] axes, respectively. (d) These correlations may correspond to the vectors connecting $Cd_{trig}$ and $Cd_{tet}$ sites as marked with a black arrow. (d) Two possible sets of structural correlations which could generate the observed quadrupole-like 3D-ΔPDF feature at the (1/3 , −1/3 , 1/8)-type vectors. Colored arrows indicate the directions of the correlated in-plane displacement of $Cd_{trig}$ and the out-of-plane displacement of $Cd_{tet}$ as inferred from the orientation of the quadrupole-like feature in the 3D-ΔPDF. This can be understood as a tendency of the $Cd_{tet}$ site to approach the nearest $P_{trig}$ site when a nearby $Cd_{trig}$ departs away from the $P_{trig}$ site.

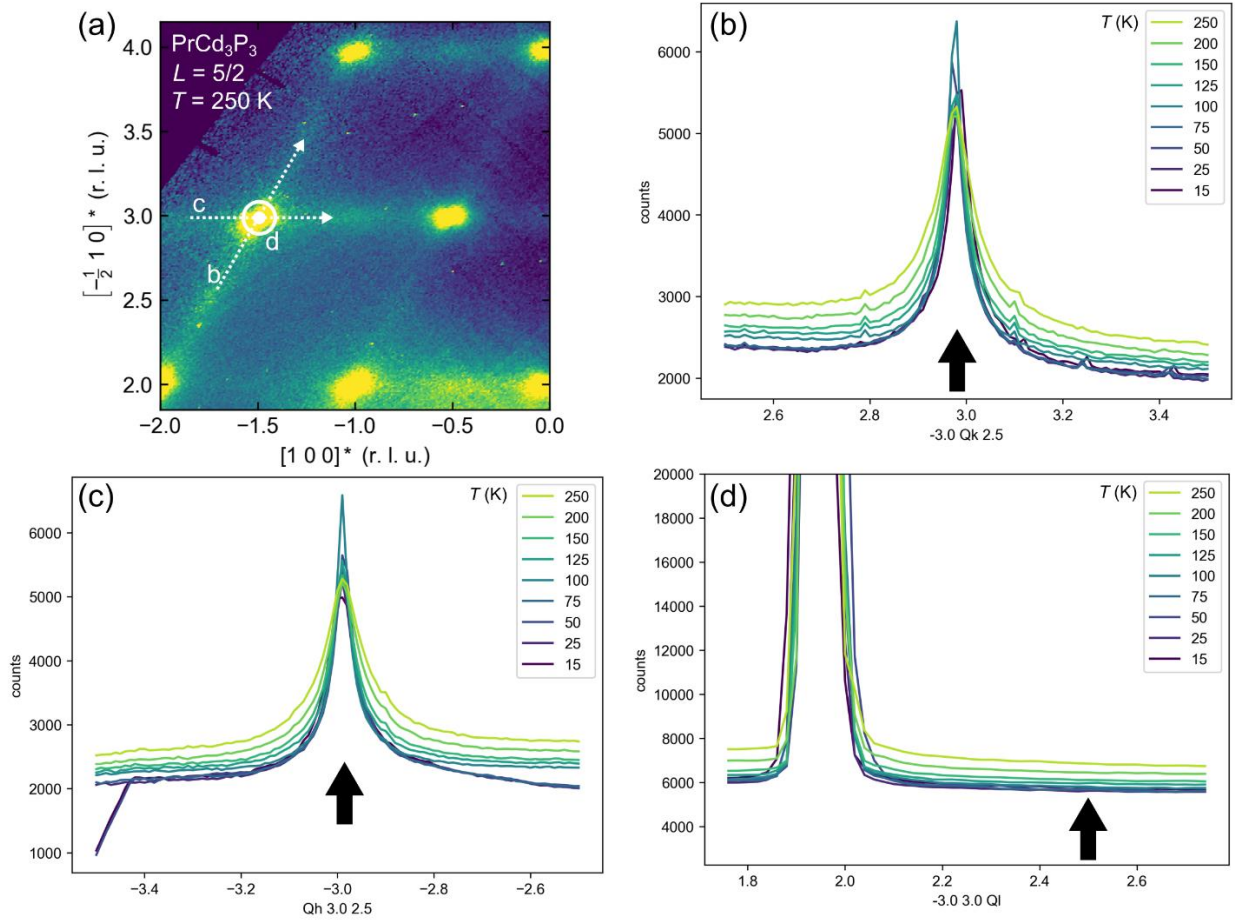

Fig. S8. (a) Line-cut directions highlighted in the ($H$, $K$, 5/2) scattering plane, centered on the tail intensity of nearby Bragg peaks in integer $L$ planes. (b) Line-cuts along $K$, (c) $H$, and (d) $L$. An arrow marks the half integer $L$ position of the scattering plane in panel (a).

## V. 3D-ΔPDF Analysis

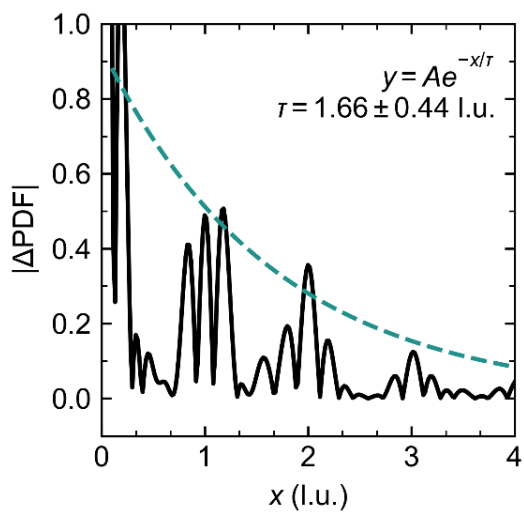

Fig. S9. Fit of |ΔPDF| along the crystallographic a direction for PrCd$_3$P$_3$ at $T$ = 300 K. An exponential decay model yields a decay constant of $\tau$ = 1.66$a$ ± 0.44$a$, indicating that the signal decays significantly within 4−5 unit cells.

VI. Forward Monte Carlo Modeling

A. Dual triangular lattice antiferromagnetic Ising model

We employed the software DISCUS to perform forward Monte Carlo modeling based on the dual triangular lattice with Ising degrees of freedom on each site. A 100 × 100 × 1 supercell was initialized with an equal population of two distinct atom types representing the pseudospins. During each run, the structure was allowed to relax via random swapping of atom types between two sites along directions determined by correlation vectors specified by the user. The move is accepted if the energy is decreased, with a probability proportional to the pseudotemperature ("*temp*").

Occupational correlation vectors were specified for first- second- and third-nearest neighbors, using a fixed energy $-J_1$, $-J_2$, and $-J_3$. A pseudotemperature = 1.0 was used for all Monte Carlo simulations. The stopping criteria was defined as 500 times the number of sites in the supercell. After the Monte Carlo simulation thermalized, the Ising model was mapped back to the dimer model to generate a corresponding honeycomb layer of Cd−P with $Cd_{trig}$ and $P_{trig}$ displaced along the bond occupied by a dimer as shown in Figure 1(f,g). For sites with overlapping dimers, one dimer was chosen at random and the other(s) discarded. The full experimental scattering volume was then simulated using the "*fourier*" section of DISCUS.

B. Three-State Potts Model

Here we provide an alternative Hamiltonian which captures the observed short-range order via a three-state Potts model on a triangular lattice. Distortions were parameterized based on the orthorhombic *Cmcm* cell from the small box modeling of the powder PDF data, though the model is not constrained by this symmetry. The Monte Carlo model assigns each $Cd_{trig}$ one of three possible distortions away from a nearest neighbor P site and was informed by chemical bonding considerations, including an energy penalty for creating under-bonded or over-bonded P sites. This places constraints on the relative distortion directions of the first coordination shell of local Cd atoms around each P. Due the triangular geometry of the network, these constraints are direction-dependent. The following Hamiltonian, derived from the three-state Potts model, captures the resulting correlations along the three directions connecting Cd sites:

$$\widehat{H}_1 = J_1 \sum_{n.n.} |e_{ijk}| \quad (1)$$

where the sum is performed over nearest neighbor (*n.n.*) sites only. The index $i$ corresponds to the crystallographic directions ⟨0, 1, 0⟩, ⟨1, 1, 0⟩, and ⟨1, 0, 0⟩, respectively, for $i$ = 1, 2, 3. The indices $j$ and $k$, which also take the values $j, k$ = {1, 2, 3}, encode the distortion directions A, B, and C on each of the two neighboring sites, as defined in Figure 1(d). When $J_1 > 0$ this Hamiltonian enforces an energy penalty for neighboring pairs which leave a phosporous atom either under- or over-bonded.

To capture the translational symmetry breaking that underlies the observed half-integer peak in the diffuse scattering along *H* and *K*, a second term promoting the formation of alternating stripes was necessary. This term $\widehat{H}_2$ also follows the same direction-dependent chemical bonding penalties, albeit perpendicular to the correlations promoted by $\widehat{H}_1$. $\widehat{H}_2$ then takes the form

$$\widehat{H}_2 = J_1 \sum_{n.n.} \delta_{ik}(1 - \delta_{ij}) \quad (2)$$

where $|J_2| < |J_1|$. The $J_2 < 0$ regime favors the formation of a Herringbone pattern.

A 100 × 100 × 1 supercell was initialized with a random distribution and equal population of each of the three $Cd_{trig}$ displacements. The magnitude of these displacements was fixed during the Monte Carlo simulation, and corresponded to the displacement obtained from the refinement of our fit to the powder PDF data. Nearest neighbor Cd sites were allowed to switch displacement directions during the simulation. The optimized model used interaction energies of $J_1/k_B T$ = 1.5 and $J_2/J_1$ = −0.4 The resulting supercell generated from this forward Monte Carlo model features one-dimensional CdP chain correlations over short length scales (3–5 unit cells), nearly identical to the results obtained from the Ising/dimer model, though there are qualitative differences in the patterns of diffuse scattering expected in the (*H*, *K*)-plane as shown in Figure S11.

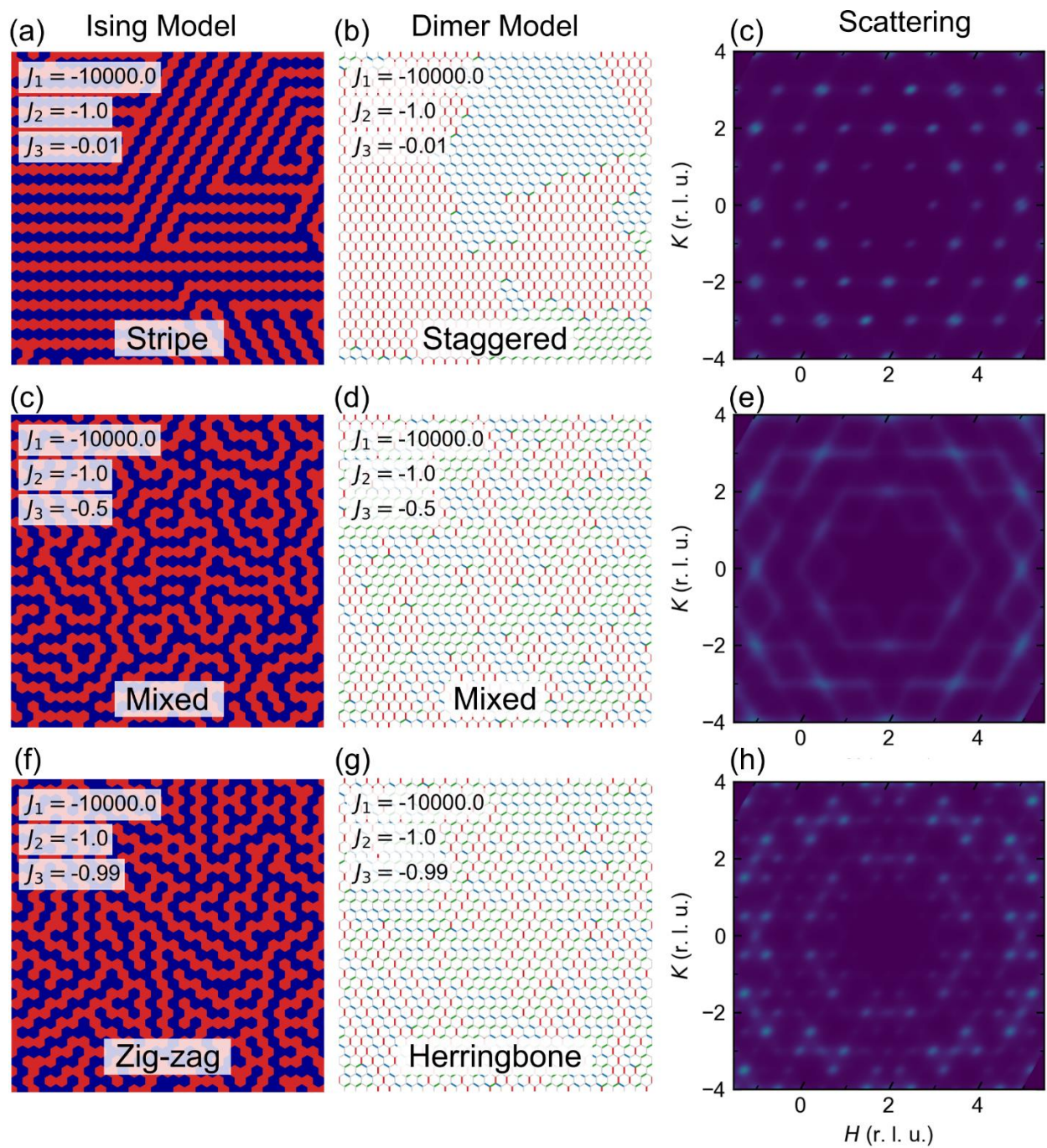

Fig. S10. Results from the forward Monte Carlo modeling. (a,e,f) The relaxed ground state of the Ising model obtained from the Monte Carlo simulation, with the interaction parameters printed in the inset. (b,d,g) The corresponding dimer state, and (c,e,h) the calculated diffuse scattering in the ($H$, $K$) plane.

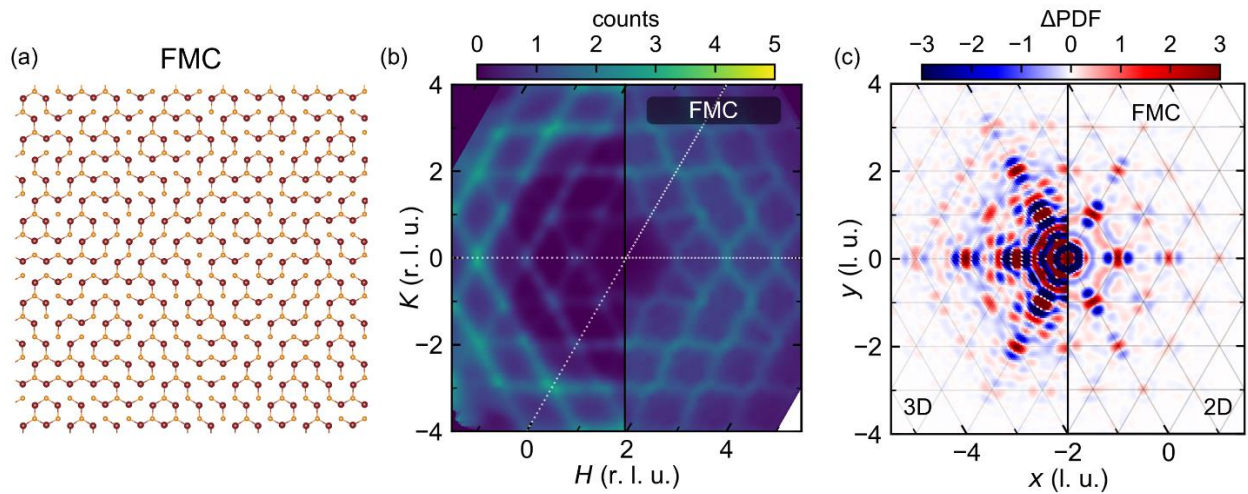

Fig. S11. (a) An example of the thermalized ground state resulting from the Forward Monte Carlo (FMC) employing a three-state Potts model, revealing short-range charge stripes composed of dimerized Cd-P pairs. (b) Experimental diffuse scattering in the ($H$, $K$, 5/2) scattering plane for PrCd$_3$P$_3$ at $T$ = 300 K isolated by a modified punch-and-fill method, and the calculated diffuse scattering from the thermalized ground state of the Monte Carlo model. (c) The 3D-$\Delta$PDF for PrCd$_3$P$_3$ extracted from the diffuse scattering at $T$ = 300 K, and the 2D-$\Delta$PDF resulting from the Monte Carlo model.

## VII. INELASTIC NEUTRON SCATTERING

To probe the crystalline electric field (CEF) about the rare earth ions in the LnCd3P3 lattice, and its potential coupling to a symmetry lowering distortion within the Cd3P3 blocks, inelastic neutron scattering (INS) data were collected at SEQUOIA (BL-17) and HYSPEC (BL-14B) direct-geometry time-of-flight chopper spectrometers at the Spallation Neutron Source (SNS) at Oak Ridge National Laboratory (ORNL). PrCd$_3$P$_3$ was chosen for this study due to its non-Kramers Pr$^{3+}$ ion. Specifically, Pr$^{3+}$ selects a non-Kramers $J = 4$ magnetic ground state, and the nine-fold degeneracy of this state is split via coupling to the CEF generated by the surrounding ligands and nearby cations. Multiplets in this state are not protected by time-reversal symmetry, and, due to their sensitivity to local structural distortions via the CEF, non-Kramers doublets in the CEF level scheme are a useful means of probing local symmetry breaking.

Using the ideal $P6_3/mmc$ structure and structural parameters determined via our x-ray scattering data, a point charge model was used to determine the Steven's parameters and eigenvalues/eigenvectors for the $J = 4$ multiplet splitting from the CEF Hamiltonian. The results are shown in Tables S2 and S3. The point charge model predicts a singlet ground state and a series of excited doublet and singlet states. The first exited state is predicted to be a doublet, and this is important because Pr$^{3+}$ is a non-Kramers ion. This doublet is not protected by time-reversal symmetry, meaning it can be split by the local structural distortions native to the Cd$_3$P$_3$ layers. If they couple to the CEF seen by the Pr$^{3+}$ ion in an appreciable way, then the first excited doublet should split.

Experimental data shown in Fig. S12(a) show scattering collected on the SEQUOIA spectrometer and a series of intramultiplet excitations. These excitations were then fit to a level scheme using the PyCrystalField software as described in the Methods section of the main text. The resulting fits to the data, the experimentally determined Steven's parameters, and the resulting eigenvectors and eigenvalues are shown in Fig. S12(a), Table S2, and Table S4 respectively. The optimized solution reveals a singlet ground state and a first excited state doublet, in agreement with the point-charge calculations (though the energy of the observed first doublet is lower relative to the point charge model).

The relatively well-isolated non-Kramer's doublet at ≈ 6.5 meV is then further investigated via high-resolution scattering measurements collected on the HYSPEC spectrometer shown in Fig. S12(b). Inspection of the doublet reveals that the two levels in the doublet are indeed slightly split by ≈0.4 meV at 1.8 K. Pr$^{3+}$ ions have a singlet ground state and the splitting persists to $T > 50$ K, which rules out a magnetic (mean field) origin to this splitting. Instead the splitting should be driven via structural symmetry lowering in the CEF felt by the Pr$^{3+}$ ions.

For a structural origin, there are two scenarios. The first is that there are two unique Pr$^{3+}$ sites in the lattice with slightly different local environments, and the second is that CEF around the Pr$^{3+}$ ion has had its symmetry lowered in an asymmetric way, splitting the non-Kramers doublet for all $Pr^{3+}$ sites. One way of discriminating between these two scenarios is to apply a magnetic field, as shown in Fig. S12(c). The applied field increases the splitting of the doublet roughly linearly and only two peaks persist (*i.e.*, instead of a quartet of two Zeeman-split peaks). This can be interpreted as a single environment with either a Zeeman term contributing to the CEF-splitting term in the Hamiltonian, or, as a magnetic field coupling to the lattice directly and modifying the relative strength of the asymmetric CEF field. The linear splitting under field shown in Fig. S12(d) may imply the latter possibility; however the density of field points is too low to draw any conclusions. Regardless, these data demonstrate that the $Pr^{3+}$ ions feel the distortion of the lattice driven via the Cd-P blocks via their CEF environment. This strongly supports the hypothesis that the structural and magnetic channels of frustration can be coupled in metallic samples.

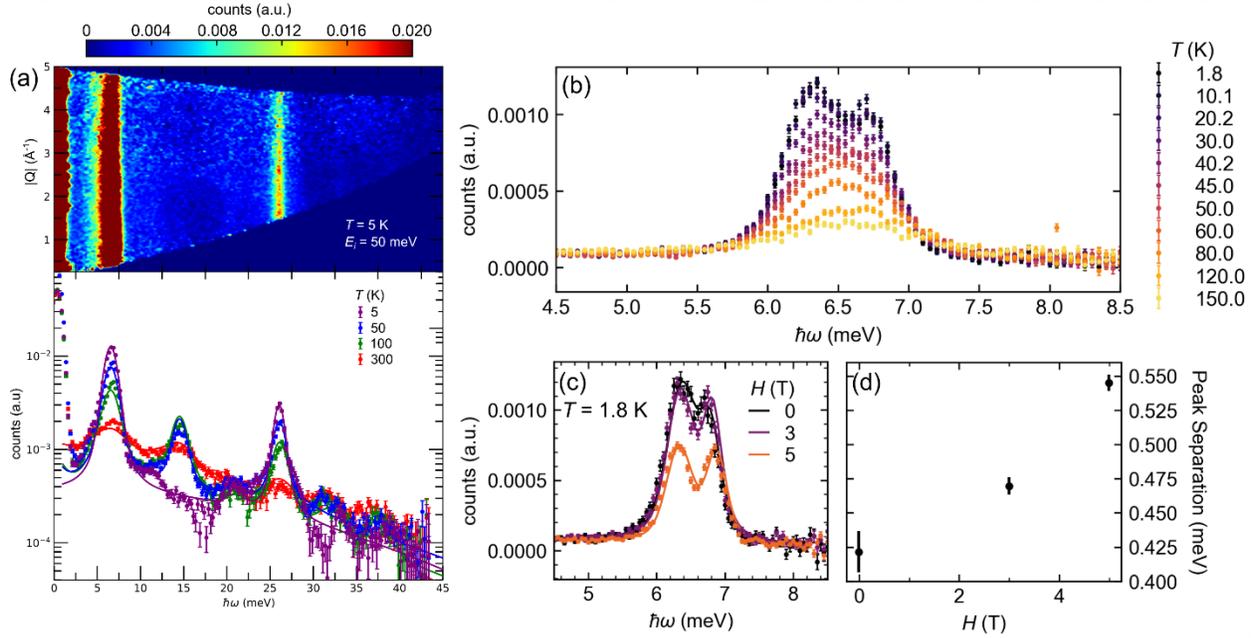

Fig. S12. (a) Powder-averaged inelastic neutron scattering intensity map of polycrystalline PrCd$_3$P$_3$ measured on SEQUOIA at $T$ = 5 K using an incident energy $E_i$ = 50 meV. The corresponding $Q$-cuts at $T$ = 5, 50, 100, and 300 K have an integration range of $Q$ = [0, 3] Å$^{-1}$. Solid lines indicate a least squares fit obtained from a point charge model with a reduced $\chi^2$ = 4.44 for $T$ = 5 K. (b) Temperature-dependent inelastic neutron scattering data for PrCd3P3 collected on HYSPEC using an incident energy $E_i$ = 9.5 meV at zero field, and (c) field-dependent data at $T$ = 1.5 K. Solid lines indicate peak fits using two pseudo-Voigt lineshapes, where the Lorentzian components of the two peaks were constrained to be equal. Points represent integrated intensity and error bars represent one standard deviation in panels (a–c). (d) The peak splitting as a function of field. Points represent parameter values obtained from non-linear least-squares fitting, and error bars indicate the 1$\sigma$ uncertainties from the fit.

TABLE S2. Determination of the Stevens parameters for PrCd$_3$P$_3$ using a point charge model based on the structure refined from powder X-ray diffraction, and those after refinement against the experimental inelastic neutron scattering (INS) data.

| $B_l^m$ | m | l | Point Charge Model | Least Squares Fit of INS |
|---|---|---|---|---|
| $B_2^0$ | 0 | 2 | 8.6066E-1 | 7.890(4)E-1 |
| $B_2^1$ | 1 | 2 | -2E-8 | 4.055(3)E-3 |
| $B_2^2$ | 2 | 2 | 0 | 2.585(4)E-1 |
| $B_4^0$ | 0 | 4 | 1.3450E-2 | 1.169(8)E-4 |
| $B_4^1$ | 1 | 4 | 0 | 2.511(4)E-3 |
| $B_4^2$ | 2 | 4 | 0 | 9.202(5)E-4 |
| $B_4^3$ | 3 | 4 | -4.19E-1 | -4.088(3) |
| $B_4^4$ | 4 | 4 | 0 | 1.997(9)E-1 |
| $B_6^0$ | 0 | 6 | 1.012E-4 | 1.163(1) |
| $B_6^1$ | 1 | 6 | 0 | -1.130(8)E-3 |
| $B_6^2$ | 2 | 6 | 0 | 8.410(0)E-3 |
| $B_6^3$ | 3 | 6 | 5.3062E-4 | 3.637(9)E-2 |
| $B_6^4$ | 4 | 6 | 0 | -1.193(0)E-4 |
| $B_6^5$ | 5 | 6 | 0 | 4.408(9)E-2 |
| $B_6^6$ | 6 | 6 | 9.1593E-4 | 1.217(1)E-1 |

TABLE S3. Energy eigenvalues (in meV) and eigenvectors (in $|J = 4, m_J\rangle$ basis) of $\hat{H}_{\text{CEF}}$ for PrCd$_3$P$_3$ employing the Stevens parameters calculated from a point charge model, as summarized in Table S2.

| $E$ (meV) | $\|-4\rangle$ | $\|-3\rangle$ | $\|-2\rangle$ | $\|-1\rangle$ | $\|0\rangle$ | $\|1\rangle$ | $\|2\rangle$ | $\|3\rangle$ | $\|4\rangle$ |
|---|---|---|---|---|---|---|---|---|---|
| 0.0 | 0 | -0.537 | 0 | 0 | 0.65 | 0 | 0 | 0.537 | 0 |
| 18.51665 | -0.205 | 0 | -0.444 | 0.511 | 0 | 0.511 | 0.444 | 0 | 0.205 |
| 18.51665 | 0.205 | 0 | -0.444 | -0.511 | 0 | 0.511 | -0.444 | 0 | 0.205 |
| 32.27336 | 0 | -0.707 | 0 | 0 | 0 | 0 | 0 | -0.707 | 0 |
| 37.50751 | 0.196 | 0 | 0.55 | -0.399 | 0 | 0.399 | 0.55 | 0 | 0.196 |
| 37.50751 | 0.196 | 0 | -0.55 | -0.399 | 0 | -0.399 | 0.55 | 0 | -0.196 |
| 65.47709 | 0 | 0.46 | 0 | 0 | 0.76 | 0 | 0 | -0.46 | 0 |
| 88.85201 | -0.648 | 0 | 0.026 | -0.283 | 0 | 0.283 | 0.026 | 0 | -0.648 |
| 88.85201 | -0.648 | 0 | -0.026 | -0.283 | 0 | -0.283 | 0.026 | 0 | 0.648 |

TABLE S4. Energy eigenvalues (in meV) and eigenvectors (in $|J = 4, m_J\rangle$ basis) of $\hat{H}_{\text{CEF}}$ for PrCd$_3$P$_3$ employing the Stevens parameters refined from a fit to inelastic neutron scattering data, as summarized in Table S2.

| $E$ (meV) | $\|-4\rangle$ | $\|-3\rangle$ | $\|-2\rangle$ | $\|-1\rangle$ | $\|0\rangle$ | $\|1\rangle$ | $\|2\rangle$ | $\|3\rangle$ | $\|4\rangle$ |
|---|---|---|---|---|---|---|---|---|---|
| 0.0 | 0 | -0.418 | 0 | 0 | -0.807 | 0 | 0 | 0.418 | 0 |
| 6.5496 | 0.363 | 0 | 0 | 0.894 | 0 | 0 | -0.261 | 0 | 0 |
| 6.5496 | 0 | 0 | -0.261 | 0 | 0 | -0.894 | 0 | 0 | 0.363 |
| 21.0765 | 0 | 0 | -0.965 | 0 | 0 | 0.246 | 0 | 0 | -0.088 |
| 21.0765 | -0.088 | 0 | 0 | -0.246 | 0 | 0 | -0.965 | 0 | 0 |
| 26.0607 | 0 | -0.707 | 0 | 0 | 0 | 0 | 0 | -0.707 | 0 |
| 32.9076 | 0 | 0.571 | 0 | 0 | -0.591 | 0 | 0 | -0.571 | 0 |
| 59.1415 | 0 | 0 | -0.011 | 0 | 0 | -0.373 | 0 | 0 | -0.928 |
| 59.1415 | -0.928 | 0 | 0 | 0.373 | 0 | 0 | -0.011 | 0 | 0 |

VIII. Doping Dependence

The impact of carrier substitution was explored via resistivity measurements shown in Fig. S13. Data are shown for the parent PrCd$_3$P$_3$ and nominally $x = 0.05$ and $x = 0.10$ Pr$_{1-x}$Sr$_x$Cd$_3$P$_3$ doped samples. The undoped PrCd$_3$P$_3$ crystal shows an extrinsic, lightly-doped behavior that metallizes as holes are introduced via Sr substitution [Fig. S13(a)]. At low temperature, a weak upturn remains, suggestive of potential disorder-induced localization effects.

Hall effect measurements confirm the presence of hole-like carriers as indicated by the positive sign [Fig. S13(b)]. Fits of the Hall resistivity to a single-band model yield a carrier concentration on the order of $4 \times 10^{18}$ cm$^{-3}$. We stress here that the dopant concentrations $x$ are only nominal amounts from the starting crystal growth and are likely larger than those incorporated into the crystals.

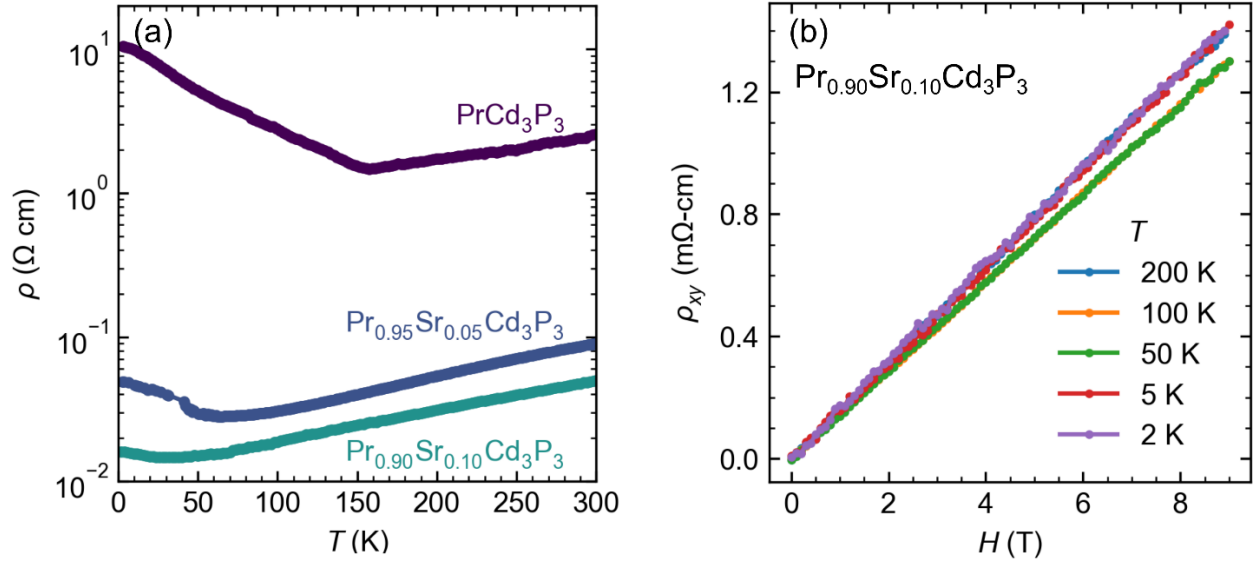

Fig. S13. (a) Doping dependence of the temperature-dependent resistivity of Pr$_{1-x}$Sr$_x$Cd$_3$P$_3$ for $x$ = 0, 0.05, and 0.10. (b) Temperature dependence of the Hall resistivity.